# Re-evaluation of the comparative effectiveness of bootstrap-based optimism correction methods in the development of multivariable clinical prediction models

Katsuhiro Iba, ME
Department of Statistical Science, School of Multidisciplinary Sciences, The Graduate University for Advanced Studies, Tokyo, Japan
Office of Biostatistics, Department of Biometrics, Headquarters of Clinical Development, Otsuka Pharmaceutical Co., Ltd., Tokyo, Japan

Tomohiro Shinozaki, PhD
Department of Information and Computer Technology, Faculty of Engineering, Tokyo University of Science, Tokyo, Japan

Kazushi Maruo, PhD
Department of Biostatistics, Faculty of Medicine, University of Tsukuba, Ibaraki, Japan

Hisashi Noma, PhD*
Department of Data Science, The Institute of Statistical Mathematics, Tokyo, Japan
ORCID: http://orcid.org/0000-0002-2520-9949

*Corresponding author: Hisashi Noma
 Department of Data Science, The Institute of Statistical Mathematics
 10-3 Midori-cho, Tachikawa, Tokyo 190-8562, Japan
 TEL: +81-50-5533-8440
 e-mail: noma@ism.ac.jp

**Abstract**

**Background:** Multivariable prediction models are important statistical tools for providing synthetic diagnosis and prognostic algorithms based on patients' multiple characteristics. Their apparent measures for predictive accuracy usually have overestimation biases (known as 'optimism') relative to the actual performances for external populations. Existing statistical evidence and guidelines suggest that three bootstrap-based bias correction methods are preferable in practice, namely Harrell's bias correction and the .632 and .632+ estimators. Although Harrell's method has been widely adopted in clinical studies, simulation-based evidence indicates that the .632+ estimator may perform better than the other two methods. However, these methods' actual comparative effectiveness is still unclear due to limited numerical evidence.

**Methods:** We conducted extensive simulation studies to compare the effectiveness of these three bootstrapping methods, particularly using various model building strategies: conventional logistic regression, stepwise variable selections, Firth's penalized likelihood method, ridge, lasso, and elastic-net regression. We generated the simulation data based on the GUSTO-I trial dataset and considered how event per variable, event fraction, number of candidate predictors, and the regression coefficients of the predictors impacted the performances. The internal validity of $C$-statistics was evaluated.

**Results:** Under relatively large sample settings (roughly, events per variable $\geq 10$), the three bootstrap-based methods were comparable and performed well. However, all three methods had biases under small sample settings, and the directions and sizes of biases were inconsistent. In general, Harrell's and .632 methods had overestimation biases when event fraction become lager. Besides, .632+ method had a slight underestimation bias when event fraction was very small. Although the bias of the .632+ estimator was relatively small, its RMSE was comparable or sometimes larger than those of the other two methods, especially for the regularized estimation methods.

**Conclusions:** In general, the three bootstrap estimators were comparable, but the .632+ estimator performed relatively well under small sample settings. However, the .632+ estimator can perform worse under small sample settings when the regularized estimation methods are adopted.

**Background**

Multivariable prediction models have been important statistical tools to provide synthetic diagnosis and prognostic algorithms based on the characteristics of multiple patients [1]. A multivariable prediction model is usually constructed using an adequate regression model (e.g., a logistic regression model for a binary outcome) based on a series of representative patients, but it is well known that their "apparent" predictive performances such as discrimination and calibration calculated in the derivation dataset are better than the actual performance for an external population [2]. This bias is known as "optimism" in prediction models. Practical guidelines (e.g., the TRIPOD statements [2, 3]) recommend optimism adjustment using principled internal validation methods, e.g., split-sample, cross-validation (CV), and bootstrap-based corrections. Among these validation methods, split-sample analysis is known to provide a relatively imprecise estimate, and CV is not suitable for some performance measures [4]. Thus, bootstrap-based methods are generally recommended because they provide stable estimates for performance measures with low biases [2, 4].

In terms of bootstrapping approaches, there are three effective methods to correct for optimism, specifically Harrell's bias correction and the .632 and .632+ estimators [1, 5, 6]. In clinical studies, Harrell's bias correction has been conventionally applied for internal validation, while the .632 and .632+ estimators are rarely seen in practice. These three estimators are derived from different concepts, and may exhibit different performances under realistic situations. Several simulation studies have been conducted to assess their comparative effectiveness. Steyerberg et al. [4] compared these estimators for ordinary logistic regression models with the maximum likelihood (ML) estimate, and reported that their performances were not too different under relatively large sample settings. Further, Mondol and Rahman [7] conducted similar simulation studies to assess their performances under rare event settings. They also considered Firth's penalized likelihood method [8, 9] for estimating regression coefficients, and concluded that the .632+ estimator performed especially well. Several other modern effective estimation methods have been widely adopted in clinical studies. Representative approaches include regularized estimation methods such as ridge [10], lasso (least absolute shrinkage and selection operator) [11], and elastic-net [12]. Also, conventional stepwise variable selections are still widely adopted in current practice [13]. Note that the previous studies of these various estimation methods did not assess the comparative effectiveness of the bootstrapping estimators. Also, their simulation settings were not very comprehensive because their simulations were conducted to assess various methods and outcome measures, and comparisons of the bootstrap estimators comprised only part of their analyses.

In this article, we conducted extensive simulation studies to provide statistical evidence concerning the comparative effectiveness of the bootstrapping estimators, as well as to provide recommendations for their practical use. In particular, we evaluated these methods using multivariable prediction models that were developed with various model building strategies: conventional logistic regression (ML), stepwise variable selection, Firth's penalized likelihood, ridge, lasso, and elastic-net. We considered extensive simulation settings based on a real-world clinical study data, the GUSTO-I trial [14, 15]. Note that we particularly focused on the evaluation of the $C$-statistic [16] in this article, since it is the most popular discriminant measure in clinical prediction models, because the simulation data should be too rich for the extensive studies; the generalizations to other performance measures would be discussed in the Discussion section.

**Methods**

*Model-building strategies for multivariable prediction models*

***Logistic regression model for a multivariable prediction model***

At first, we briefly introduce the estimating methods for the multivariable prediction models. We consider constructing a multivariable prediction model for a binary outcome. A logistic regression model is widely used as a regression-based prediction model [2, 17]. Let $y_i$ $(i = 1,2, ..., n)$ be a binary outcome variable ($y_i = 1$: event occurring, or 0: not occurring) and let $\boldsymbol{x}_i = (x_{i1}, x_{i2}, ..., x_{ip})^T$ $(i = 1,2, ..., n)$ be $p$ predictor variables for an individual $i$. Based on the logistic regression model, the probability of event occurrence $\pi_i = \Pr(y_i = 1|\boldsymbol{x}_i)$ is modelled as

$$\pi_i = \frac{\exp(\beta_0 + \beta_1 x_{i1} + \beta_2 x_{i2} + \cdots + \beta_p x_{ip})}{1 + \exp(\beta_0 + \beta_1 x_{i1} + \beta_2 x_{i2} + \cdots + \beta_p x_{ip})}$$

where $\boldsymbol{\beta} = (\beta_0, \beta_1, ..., \beta_p)^T$ are the regression coefficients containing the intercept. Plugging an appropriate estimate $\widehat{\boldsymbol{\beta}}$ of $\boldsymbol{\beta}$ into the above model, the estimated probability $\hat{\pi}_i$ $(i = 1,2, ..., n)$ is defined as the risk score of individual patients, and this score is used as the criterion score to determine the predicted outcome.

The ML estimates of $\widehat{\boldsymbol{\beta}}_{ML}$ for $\boldsymbol{\beta}$ are obtained by maximizing the log likelihood function

$$l(\boldsymbol{\beta}) = \sum_{i=1}^{n} y_i \log \pi_i + (1 - y_i) \log(1 - \pi_i)$$

The ordinary ML estimation can be easily implementable by standard statistical software and has been widely adopted in practice [2, 17]. However, the ML-based modelling strategy is known to have several finite sample problems when applied to a small or sparse dataset. In particular, the regression coefficient estimators may be associated with overestimation bias [18, 19] and can produce optimistic estimates for discriminant measures. Also, the model estimation can become unstable when the predictor effects are large or sparse [9, 20]. The estimated probabilities tend to be close to 0 or 1, and can perfectly discriminate between events and non-events. Both properties disappear with increasing events per variables (EPVs), defined as the ratio of the number of events to the number of predictor variables (*p*) of the prediction model. EPV has been formally adopted as a sample size criterion for model development, and in particular, EPV ≥ 10 is a widely adopted criterion [21]; recent studies showed that the validity of this criterion depended on case-by-case conditions [4, 13, 22]. As noted below, the following shrinkage estimation methods can moderate these problems.

For variable selection, some automated mathematical algorithms are available. The representative example is stepwise selection [23]. Forward and backward strategies can be adopted, with the latter generally recommended for the development of prediction models [23]. For the stopping rule, the significance level criterion (a conventional threshold is P < .050), Akaike Information Criterion (AIC) [24], and Bayesian Information Criterion (BIC) [25] can be adopted.

*Firth's logistic regression*

In order to reduce the first-order finite sample bias of the ML estimator, Firth's penalized likelihood logistic regression was developed [8]. In Firth's method, the regression coefficients are estimated using a modified log likelihood function with a penalty term,

$$l(\boldsymbol{\beta}) + \frac{1}{2}\log|I(\boldsymbol{\beta})|$$

where $I(\boldsymbol{\beta})$ is the Fisher's information matrix [9]. By including the penalty term, the modified log-likelihood could have a maximum even under (quasi-)complete separation, which causes infinite value of ML estimates [9]. Moreover, the penalty term would shrink an estimate towards 0 based on information contained in each predictor variable. As a result, the regression coefficients estimate becomes stable even for small or sparse datasets [9]. However, if the Fisher information is degenerated, the resultant estimator can be instable.

*Ridge regression*

Ridge regression was originally proposed to deal with collinearity among predictor variables; the ML estimate can be unstable if strong collinearity exists [10]. The ridge estimation is a shrinkage estimation performed by the following penalized log likelihood function:

$$l(\boldsymbol{\beta}) - \lambda \sum_{j=1}^{p} \beta_j^2$$

where $\lambda \ (> 0)$ is a turning parameter to control the degree of shrinkage. Note that the predictors are usually standardized in advance to have mean 0 and variance 1. The penalty term for ridge regression is the sum of squares of the regression coefficients, and the resultant regression coefficients estimate is shrunk towards zero and thereby can reduce overfitting [26]. Several methods, such as CV, were proposed for selection of the tuning parameter $\lambda$ [17, 26, 27].

*Lasso regression*

Lasso regression is another well-known shrinkage estimation method. The penalized log likelihood function of lasso [11] is defined as

$$l(\boldsymbol{\beta}) - \lambda \sum_{j=1}^{p} |\beta_j|$$

where $\lambda$ is a positive turning parameter similar to that in ridge regression. The penalty term for lasso is the sum of the absolute value of regression coefficients. Note that the predictors are usually standardized in advance to have mean 0 and variance 1. In contrast to ridge regression, lasso can shrink some regression coefficients to be exactly 0; therefore, lasso can perform shrinkage estimation and variable selection simultaneously. However, lasso is known to perform poorly if highly correlated predictors exist [12].

*Elastic-net regression*

Elastic-net [12] is an effective method of shrinkage regression that was proposed to overcome the disadvantage of lasso while still retaining the variable selection property. The penalty of elastic-net is constructed by combining the penalties of lasso and ridge:

$$l(\boldsymbol{\beta}) - \lambda \left\{ (1-\alpha) \sum_{j=1}^{p} \beta_j^2 + \alpha \sum_{j=1}^{p} |\beta_j| \right\}$$

where $\lambda \ (>0)$ is the turning parameter to determine the degree of shrinkage and $\alpha \ (0 \leq \alpha \leq 1)$ is the weight of the lasso and ridge penalties. Even when there are strong correlations among the predictor variables, the strongly correlated predictor variables can remain together for elastic-net regression. Also, similar to lasso, some coefficients shrink to exactly 0 when the modified log likelihood function is used [12]. The predictors should also be standardized in advance to have mean 0 and variance 1.

*Software packages*

In numerical analyses in the following sections, all of the above methods were performed in R ver. 3.5.1 [28]. The ordinary logistic regression was fitted by the `glm` function. The stepwise variable selections were performed using the `stats` and `logistf` packages [29]. Firth's logistic regression was also conducted using the `logistf` package [29]. The ridge, lasso, and elastic-net regressions were performed using the `glmnet` package [30]; the turning parameters were consistently determined using the 10-fold CV of deviance.

*C-statistic and the optimism-correction methods based on bootstrapping*

**C-statistic**

Secondly, we review the internal validation methods used in the numerical studies. We focused especially on the *C*-statistic in our numerical evaluations, as it is most frequently used in clinical studies as an summary measure of the discrimination of prediction models [2]. The *C*-statistic is defined as the empirical probability that a randomly selected patient who has experienced an event has a larger risk score than a patient who has not experienced the event [16]. The *C*-statistic also corresponds to the area under the curve (AUC) of the empirical receiver operating characteristic (ROC) curve for the risk score. The *C*-statistic ranges from 0.5 to 1.0, with larger values corresponding to superior discriminant performance.

**Harrell's bias correction method**

The most widely applied method for bootstrapping optimism correction is Harrell's bias correction [1], which is obtained by the conventional bootstrap bias correction method [4]. The algorithm is summarized as follows:

- Let $\theta_{app}$ be the apparent predictive performance estimate for the original population.
- Generate $B$ bootstrap samples by resampling with replacement from the original population.
- Construct a prediction model for each bootstrap sample, and calculate the predictive performance estimate for it. We denote the $B$ bootstrap estimates as

$\theta_{1,boot}, \theta_{2,boot}, \cdots, \theta_{B,boot}$.

- Using the *B* prediction models constructed from the bootstrap samples, calculate the predictive performance estimates for the original population: $\theta_{1,orig}, \theta_{2,orig}, \cdots, \theta_{B,orig}$.

- The bootstrap estimate of optimism is obtained as

$$\Lambda = \frac{1}{B}\sum_{b=1}^{B}(\theta_{b,boot} - \theta_{b,orig})$$

Subtracting the estimate of optimism from the apparent performance, the bias corrected predictive performance estimate is obtained as $\theta_{app} - \Lambda$.

The bias correction estimator is calculable by a relatively simple algorithm, and some numerical evidence has shown that it performs well under realistic situations [4]. Therefore, this algorithm is currently adopted in most clinical prediction model studies that conduct bootstrap-based internal validations. However, a certain proportion of patients in the original population (on average, 63.2%) should be overlapped in the bootstrap sample. The overlap may cause overestimation of the predictive performance [31], and several alternative estimators have therefore been proposed, as follows.

*Efron's .632 method*

Efron's .632 method [5] was proposed as a bias-corrected estimator that considers overlapped samples. Among the *B* bootstrap samples, we consider the "external" samples that are not included in the bootstrap samples to be 'test' datasets for the developed *B* prediction models. Then, we calculate the predictive performance estimates for the external samples by the developed *B* prediction models $\theta_{1,out}, \theta_{2,out}, \cdots, \theta_{B,out}$, and we denote the average measure as $\theta_{out} = \sum_{b=1}^{B}\theta_{b,out}/B$. Thereafter, the .632 estimator is defined as a weighted average of the predictive performance estimate in the original sample $\theta_{app}$ and the external sample estimate $\theta_{out}$:

$$\theta_{.632} = 0.368 \times \theta_{app} + 0.632 \times \theta_{out}$$

The weight .632 derives from the approximate proportion of subjects included in a bootstrap sample. Since the subjects that are included and not included in a bootstrap sample are independent, the .632 estimator can be interpreted as an extension of CV [4, 7]. However, the .632 estimator is associated with overestimation bias under highly overfit situations, when the apparent estimator $\theta_{app}$ has a large bias [6].

*Efron's .632+ method*

Efron and Tibshirani [6] proposed the .632+ estimator to address the problem of the .632 estimator by taking into account the amount of overfitting. They define the relative overfitting rate $R$ as

$$R = \frac{\theta_{out} - \theta_{app}}{\gamma - \theta_{app}}$$

$\gamma$ corresponds to 'no information performance', which is the predictive performance measure for the original population when the outcomes are randomly permuted. In the case of the *C*-statistic, $\gamma$ should theoretically take a value of 0.50 [4]. The overfitting rate $R$ approaches 0 when there is no overfitting ($\theta_{out} = \theta_{app}$), and approaches 1 when the degree of overfitting is large. Then, the .632+ estimator [6] is defined as

$$\theta_{.632+} = (1 - w) \times \theta_{app} + w \times \theta_{out}$$

$$w = \frac{.632}{1 - .368 \times R}$$

Note that the weight $w$ ranges from .632 ($R = 0$) to 1 ($R = 1$). Hence, the .632+ estimator approaches the .632 estimator when there is no overfitting and approaches the external sample estimate $\theta_{out}$ when there is marked overfitting.

In the following numerical studies, the numbers of bootstrap resamples were consistently set to $B = 2000$. Also, for the model-building methods involving variable selections (e.g., stepwise regression) and shrinkage methods which require tuning parameter selections (ridge, lasso, and elastic-net), all estimation processes were included in the bootstrapping analyses in order to appropriately reflect their uncertainty.

**Real-data example: Applications to the GUSTO-I trial**

Since we consider the GUSTO-I trial as a model example for the simulation settings, we first illustrate the prediction model analyses for this clinical trial. The GUSTO-I dataset has been adopted by many performance evaluation studies of multivariable prediction models [4, 13, 27], and we specifically used the West region dataset [23]. GUSTO-I was a comparative clinical trial to assess four treatment strategies for acute myocardial infarction [14]. Here we adopted death within 30 days as the outcome variable (binary). The summary of the outcome and the 17 covariates are presented in Table 1. Of the 17 covariates, two variables (height and weight) are continuous, one variable (smoking) is ordinal, and remaining 14 variables are binary; age was dichotomized at 65 years old. For smoking, which is a three-category variable (current smokers, ex-smokers, and never smokers), we generated two dummy variables (ex-smokers vs. never smokers, and current smokers vs. never smokers) and used them in the analyses.

We considered two modelling strategies: (1) 8-predictor models (age >65 years, female gender, diabetes, hypotension, tachycardia, high risk, shock, and no relief of chest pain), which was adopted in several previous studies [4, 32]; and (2) 17-predictor models which included all variables presented in Table 1. The EPVs for these models were 16.9 and 7.5, respectively. For the two modelling strategies, we constructed multivariable logistic prediction models using the estimating methods in the Methods section (ML, Firth, ridge, lasso, elastic-net) and backward stepwise methods with AIC and statistical significance (P < 0.05). We also calculated the *C*-statistics and the bootstrap-based optimism-corrected estimates for these prediction models. The results are presented in Table 2.

For the 8-predictor models, the lasso and elastic-net models selected all eight variables. Also, the two backward stepwise methods selected the same six variables for the 8-predictor models; only 'diabetes' and 'no relief of chest pain' were excluded. Further, for the 17-predictor models, the lasso and elastic-net models selected the same 12 variables, while 'diabetes', 'height', 'hypertension history', 'ex-smoker', 'hypercholesterolemia', and 'family history of MI' were excluded. For the backward stepwise selections, the AIC method selected nine variables and the significance-based method selected seven variables; see Table 2 for the selected variables.

As a whole, comparison of the apparent *C*-statistics showed that those of the 17-predictor models were larger than those of the 8-predictor models. Although the apparent *C*-statistics of all the prediction models were not very different for the 8-predictor models, the *C*-statistic of the backward stepwise selection (P < 0.05) was slightly smaller than those of the others for the 17-predictors models. The optimism-corrected *C*-statistics were smaller than the apparent *C*-statistics for all models, and certain biases were discernible. However, among the three bootstrapping methods, there were no differences for any of the prediction models. Note that the optimism-corrected *C*-statistics between the 8- and 17-predictor models were not very different. These results indicate that the 17-predictor model had greater optimism, possibly due to the inclusion of noise variables.

**Simulations**

*Data generation and simulation settings*

As described in the previous section, we conducted extensive simulation studies to assess the bootstrap-based internal validation methods based on a real-world dataset, the GUSTO-I West region data. We considered a wide range of conditions with various factors that can affect predictive performance: the EPV (3, 5, 10, 20, and 40), the expected event fraction (0.5, 0.25, 0.125, and 0.0625), the number of candidate predictors (eight variables, as specified in the previous studies, and all 17 variables), and the regression coefficients of the predictor variables (two scenarios, as explained below). All combinations of these settings were covered, and a total of 80 scenarios were investigated. The settings of the EPV and the event fraction were based on those used in previous studies [4, 17]. For the regression coefficients of the predictor variables (except for the intercept $\beta_0$), we considered two scenarios: one fixed to the ML estimate for the GUSTO-I dataset (scenario 1) and the other fixed to the elastic-net estimate for the GUSTO-I dataset (scenario 2); in scenario 1, all the predictors have some effect on the risk of events, while in scenario 2 some of the predictor effects are null and the others are relatively small compared with scenario 1. The intercept $\beta_0$ was set to properly adjust the event fractions. The sample size of the derivation cohort *n* was determined by (the number of candidate predictor variables $\times$ EPV) / (expected event fraction).

The predictor variables were generated as random numbers based on the parameters estimated from the GUSTO-I dataset. Three continuous variables (height, weight, and age) were generated from a multivariate normal distribution with the same mean vector and covariance matrix as in the GUSTO-I data; the age variable was dichotomized at age 65 years, similar to the analyses in the real-data example. For smoking, an ordinal variable, random numbers were generated from multinomial distribution using the same proportions as in the GUSTO-I data; this variable was converted to two dummy variables before being incorporated into the prediction models. In addition, the remaining binary variables were generated from a multivariate binomial distribution [33] using the same marginal probabilities and correlation coefficients estimated from the GUSTO-I dataset. We used the `mipfp` package [34] for generating the correlated binomial variables. The event occurrence probability $\pi_i$ $(i = 1,2,...,n)$ was determined based on the generated predictor variables $\boldsymbol{x}_i$ and the logistic regression model $\pi_i = 1/(1 + \exp(-\boldsymbol{\beta}^T \boldsymbol{x}_i))$. The outcome variable $y_i$ was generated from a Bernoulli distribution with a success probability $\pi_i$.

The actual prediction performances of the developed models were assessed by 500,000 independently generated external test samples; the empirical *C*-statistics for the test datasets, which we refer to as the 'external' *C*-statistics, were used as the estimands. The number of simulations was consistently set to 2,000 for all the scenarios. Based on the generated derivation cohort dataset, the multivariable prediction models were constructed using the seven modelling strategies (ML, Firth, ridge, lasso, elastic-net, and backward stepwise selections with AIC and P < 0.05). The *C*-statistics for the derivation cohort were estimated by the apparent, Harrell, .632, and .632+ bootstrapping methods in the derivation cohort. For the evaluation measures, biases and RMSEs (root mean squared errors) for the estimated *C*-statistics from the external *C*-statistic for the 500,000 test datasets were used.

*Simulation results*

The full set of the simulation results is provided in e-Appendix A in the Supplementary Materials because the contents were too large to present in this manuscript. In the main Tables and Figures, we present the results of scenario 2 with event fractions of 0.5 and 0.0625; as mentioned below, the overall trends were not very different, and these scenarios are representative of the all simulation results. In the small-sample settings (low EPVs and large event fractions), the simulation results for modelling strategies involving variable selections (lasso, elastic-net, and stepwise selections) occasionally dropped all of the predictors; in such cases, only an intercept remained, yielding a nonsensical prediction model. We excluded these cases from the performance evaluations because such models are not usually adopted in practice. As reference information, the frequencies with which the intercept models occurred are presented in e-Table 1 in the Supplementary Materials.

The average values of the apparent, external, and optimism-corrected $C$-statistics for 2,000 simulations are shown in Figure 1 (for event fraction = 0.5) and Figure 2 (for event fraction = 0.0625). Under event fraction = 0.5, the external $C$-statistics for the test datasets were 0.65–0.70 for EPV = 3 and 5, and around 0.72 for larger EPV. These results were similar between the 8- and 17-predictor models. Further, the external $C$-statistics for stepwise selections were relatively small in general, and the actual predictive performances of these modelling strategies were relatively worse than the other methods, as shown by previous studies [17]. Also, the external $C$-statistics for ML estimation were relatively smaller when EPV were small; those of Firth regression were a bit larger than those of ML estimation in these cases. Ridge, lasso, and elastic-net showed slightly larger external $C$-statistics compared with Firth regression. Similar trends were observed under event fraction = 0.0625, but the external $C$-statistics were relatively large, around 0.75, under EPV = 3–5. These results were caused by the total sample sizes: the event fraction of the latter scenario was smaller and the total sample sizes were larger if EPV were the same. Also, the external $C$-statistics of the 17-predictor models were smaller than those of the 8-predictor models, especially when EPV were small. The external $C$-statistics of the stepwise regression were relatively small. Also, the shrinkage estimation methods provided larger external $C$-statistics compared with ML estimation when EPV were small. Also, under small EPV settings, the external $C$-statistics of ridge, lasso, and elastic-net were a bit larger than those of Firth regression. In particular, for the 8-predictor models, the external $C$-statistics of ML estimation were relatively small when EPV were small.

In Figures 3 and 4, we present the empirical biases of the estimators of $C$-statistics derived from the external $C$-statistics. Under all settings and methods, the apparent $C$-statistics had relatively large overestimation biases. In particular, under smaller EPV settings, the biases were larger. For the same EPV settings, the biases were smaller when the event fraction was smaller (i.e., the total sample sizes were larger). For event fraction = 0.5, the biases of the apparent $C$-statistics of ML estimation were larger than those of the other methods for the 8-predictor models. The biases of the stepwise regression were the smallest, but the external $C$-statistics were also smaller, as noted above. Also, the biases of ridge, lasso, and elastic-net were comparable and were a bit smaller than those of the Firth method. For the 17-predictor models, the overall trends were similar to those of the 8-predictor models, but lasso showed slightly smaller biases than ridge and elastic-net. Generally, similar trends were observed for event fraction = 0.0625. Comparing the three bootstrapping methods under EPV $\geq$ 20, the biases of all the methods were comparable for all settings. The results for the conventional ML estimation were consistent with those of Steyerberg et al. (2001) [4], and we confirmed that similar results were obtained for the shrinkage estimation methods. Further, under small EPV settings, unbiased estimates were similarly obtained by the three bootstrapping methods for the 8-predictor models with event fraction = 0.0625, since the total sample size was relatively large, and similar trends were observed for all estimation methods under EPV $\geq$ 5. Under EPV = 3, the .632+ estimator had underestimation biases, while for the ML estimation, the underestimation bias was about 0.02. For ridge, lasso, and elastic-net, the underestimation biases were about 0.01. For the Firth regression and stepwise methods, the biases were relatively small. The Harrell and .632 estimators were comparable, and they had small overestimation biases. For the 17-predictor models, the biases of the .632+ estimator were relatively small, but in general this estimator displayed underestimation biases. The Harrell and .632 estimators had overestimation biases, and for the ML and Firth methods, the biases were larger than those of the 8-predictor models. Under event fraction = 0.5, the overestimation biases of the Harrell and .632 estimators were remarkably large under EPV = 3 and 5; under the 8-predictor models, the overestimation biases were 0.03–0.04. Although the .632+ estimator had small underestimation biases for the ML and Firth methods (approximately −0.01 under EPV = 3), mostly unbiased estimates were obtained for the ridge, lasso, and elastic-net estimators. For the stepwise selections, the AIC method provided mostly unbiased estimators, but the $P < 0.05$ criterion resulted in overestimation biases that were comparable to those of the Harrell estimator. For the 17-predictor models, similar trends were observed. The Harrell and .632 estimators had overestimation biases (about 0.02–0.04 under EPV = 3), and the two estimators were comparable for the ML, Firth, and ridge estimators. However, for the lasso, elastic-net, and stepwise (AIC) methods, the biases of the Harrell estimator were smaller. For the stepwise selection method ($P < 0.05$), the .632 estimator showed overestimation bias but the Harrell estimator was mostly unbiased. Further, the .632+ estimator had small underestimation biases for the ML, Firth, and stepwise (AIC) methods; this estimator was mostly unbiased for the ridge, lasso, elastic-net, and stepwise ($P < 0.05$) methods.

The empirical RMSEs are presented in Figures 5 and 6. Under all the settings, the apparent $C$-statistics had relatively large RMSEs, due to their overestimation biases. The RMSEs of the three bootstrapping methods were generally not very different from each other. An exception is that under event fraction = 0.5 and EPV = 3 and 5, the RMSEs of the .632+ estimators of the 8-predictor models with ridge, lasso, and elastic-net were relatively larger than those of the other two estimators. As mentioned above, under these conditions the absolute biases of the .632+ estimators were smaller, reflecting these estimators' standard errors. For the 17-predictor models, the RMSEs were comparable. Under event fraction = 0.0625, the .632+ estimators for ridge, lasso, and elastic-net had relatively large RMSEs for the 8-predictor models under EPV = 3. These results reflect the fact that under these conditions, the absolute biases of the .632+ estimators were relatively large. In addition, under the other settings with small EPV, there were many scenarios in which the biases of the .632+ estimator were relatively small, and in these cases the RMSEs were comparable with those of the Harrell and .632 estimators. These results indicate the .632+ estimator had relatively large standard errors when EPV were small.

The results described above are mostly consistent with those derived under the other settings presented in the e-Appendices of the Supplementary Materials available online.

## Discussion

Bootstrapping methods for internal validations of discriminant and calibration measures in developing multivariable prediction models have been increasingly used in recent clinical studies. The Harrell, .632, and .632+ estimators are asymptotically equivalent, but in practice they might have different properties in finite sample situations. This fact may influence the main conclusions of relevant studies, and conclusive evidence of these estimators' comparative effectiveness is needed to ensure appropriate research practices.

In this article, we conducted extensive simulation studies to assess these methods under a wide range of situations. In particular, we assessed their properties in the context of the prediction models developed by modern shrinkage methods (ridge, lasso, and elastic-net), which are becoming increasingly more popular. We also evaluated stepwise selections, which are additional standard methods of variable selection, taking into consideration the uncertainties of variable selection processes.

Conventionally, the rule-of-thumb criterion for sample size determination in prediction model studies is EPV $\geq$ 10 [21]. In our simulations, the internal validation methods generally worked well under these settings. However, several counterexamples were reported in previous studies [4, 13, 22], so this should not be an absolute criterion. Also, the external $C$-statistics of the stepwise selections were smaller than those of ML estimation under certain situations, as previously discussed [13], and variable selection methods might not be recommended in practice. Moreover, the shrinkage regression methods (ridge, lasso, elastic-net) provided larger $C$-statistics than ML estimation and Firth's method under certain settings, and were generally comparable. Further investigations are needed to assess the practical value of these methods in clinical studies.

Among the bootstrapping optimism-correction methods, we showed that the Harrell and .632 methods had upward biases at EPV = 3 and 5. The biases in these methods increased when the event fraction became larger. As mentioned in the Methods section, the overlap between the original and bootstrap samples under small sample settings could cause these biases. Therefore, these methods should be used with caution in cases of small sample settings. When the event fraction was 0.5, the .632 estimator often had a greater upward bias than the Harrell method for the shrinkage estimating methods and stepwise selections. Similarly, the .632+ estimator showed upward biases at EPV = 3 for the stepwise selection ($P < 0.05$) for the 8-predictor model. Since the .632+

estimator is constructed by a weighted average of the apparent performance and the out-of-sample performance measures, it cannot have a negative bias when the resultant prediction model has extremely low predictive performance, i.e., when the apparent *C*-statistics are around 0.5. However, if such a prediction model is obtained in practice, we should not adopt it as the final model.

Also, since the .632+ estimator was developed to overcome the problems of the .632 estimator under highly overfitted situations, the .632+ estimator is expected to have smaller overestimation bias compared with the other two methods. However, the .632+ estimator showed a slight downward bias when the event fraction was 0.0625; the relative overfitting rate was overestimated in that case, since a small number of events was discriminated well by less overfitted models. This tendency was clearly shown for the ML method, which has the strongest overfitting risk.

Although the bias of the .632+ estimator was relatively small, its RMSE was comparable or sometimes larger than those of the other two methods. Since the .632+ estimator adjusts the weights of apparent and out-of-sample performances using the relative overfitting rate, the .632+ estimator has variations due to the variability of the estimated models under small sample settings. Also, the RMSE of the .632+ estimator was particularly large in the shrinkage estimation methods; the penalty parameters were usually selected by 5- or 10-fold CV and we adopted the latter in our simulations. Since the 10-fold CV is unstable with small samples [35], the overfitting rate often has large variations. We attempted to use the leave-one-out CV instead of the 10-fold CV, and this decreased the RMSE of the .632+ estimator (see e-Table 2 in Supplementary Materials). On the other hand, the RMSE in the Harrell method became larger. These results indicate that the performances of the optimism-corrected estimators depend on the methods of penalty parameter selections.

In this paper, we conducted simulation studies based on the GUSTO-I study. We considered a wide range of settings by varying several factors to investigate detailed operating characteristics of the bootstrapping methods. A limitation of the study is that parameter settings were based on the GUSTO-I study, but we took care for covering similar settings that were considered in previous large simulation studies [4, 17]. In addition, we assessed only the *C*-statistic in this study. Other measures such as the Brier score and calibration slope can also be considered for the evaluation of optimism corrections. However, in previous simulation studies, these measures showed similar trends [4].

## Conclusions

In conclusion, under certain sample sizes (roughly, EPV $\geq$ 10), all of the internal validation methods based on bootstrapping performed well. However, under small sample settings, all the methods had biases. For the ridge, lasso, and elastic-net methods, although the bias of the .632+ estimator was relatively small, its RMSE could become larger than those of the Harrell and .632 estimators. Under small sample settings, the penalty parameter selection strategy should be carefully considered; one possibility is to adopt the leave-one-out CV instead of the 5- or 10-fold CV. For the other estimation methods, the three bootstrap estimators were comparable in general, but the .632+ estimator performed relatively well under certain settings. In addition, developments of new methods to overcome these issues are future issues to be investigated.

**Funding**

This work was supported by CREST from the Japan Science and Technology Agency (Grant number: JPMJCR1412), the Practical Research for Innovative Cancer Control (Grant number: 17ck0106266) from the Japan Agency for Medical Research and Development, and the JSPS KAKENHI (Grant numbers: JP17H01557 and JP17K19808).

**Acknowledgements**

The authors are grateful to Professor Hideitsu Hino for his helpful comments and advice.

**References**

1. Harrell FE, Lee KL, Mark DB. Multivariable prognostic models: issues in developing models, evaluating assumptions and adequacy, and measuring and reducing errors. Stat Med. 1996;15(4):361-87.

2. Moons KG, Altman DG, Reitsma JB, Ioannidis JP, Macaskill P, Steyerberg EW, et al. Transparent Reporting of a multivariable prediction model for Individual Prognosis or Diagnosis (TRIPOD): explanation and elaboration. Ann Intern Med. 2015;162(1):W1-73.

3. Collins GS, Reitsma JB, Altman DG, Moons KG. Transparent Reporting of a multivariable prediction model for Individual Prognosis or Diagnosis (TRIPOD): the TRIPOD statement. Ann Intern Med. 2015;162(1):55-63.

4. Steyerberg EW, Harrell FE, Borsboom GJJM, Eijkemans MJC, Vergouwe Y, Habbema JDF. Internal validation of predictive models. J Clin Epidemiol. 2001;54(8):774-81.

5. Efron B. Estimating the error rate of a prediction rule: improvement on cross-validation. J Am Stat Assoc. 1983;78(382):316-31.

6. Efron B, Tibshirani R. Improvements on Cross-Validation: The .632+ bootstrap method.

J Am Stat Assoc. 1997;92(438):548-60.

7. Mondol M, Rahman MS. A comparison of internal validation methods for validating predictive models for binary data with rare events. J Stat Res. 2018;51:131-44.

8. Firth D. Bias reduction of maximum likelihood estimates. Biometrika. 1993;80(1):27-38.

9. Heinze G, Schemper M. A solution to the problem of separation in logistic regression. Stat Med. 2002;21(16):2409-19.

10. Lee AH, Silvapulle MJ. Ridge estimation in logistic regression. Commun Stat Simul Comput. 1988;17(4):1231-57.

11. Tibshirani R. Regression shrinkage and selection via the lasso. J R Statist Soc B. 1996;58(1):267-88.

12. Zou H, Hastie T. Regularization and variable selection via the elastic net. J R Statist Soc B. 2005;67(2):301-20.

13. Steyerberg EW, Eijkemans MJC, Harrell FE, Habbema JDF. Prognostic modelling with logistic regression analysis: a comparison of selection and estimation methods in small data sets. Stat Med. 2000;19(8):1059-79.

14. The Gusto investigators. An international randomized trial comparing four thrombolytic strategies for acute myocardial infarction. N Engl J Med. 1993;329(10):673-82.

15. Lee KL, Woodlief LH, Topol EJ, Weaver WD, Betriu A, Col J, et al. Predictors of 30-day mortality in the era of reperfusion for acute myocardial infarction. Results from an international trial of 41,021 patients. GUSTO-I Investigators. Circulation. 1995;91(6):1659-68.

16. Meurer WJ, Tolles J. Logistic regression diagnostics: Understanding how well a model predicts outcomes. JAMA. 2017;317(10):1068-9.

17. van Smeden M, Moons KG, de Groot JA, Collins GS, Altman DG, Eijkemans MJ, et al. Sample size for binary logistic prediction models: Beyond events per variable criteria. Stat Methods Med Res. 2019;28(8):2455-74.

18. Gart JJ, Zweifel JR. On the bias of various estimators of the logit and its variance with application to quantal bioassay. Biometrika. 1967;54(1/2):181–7.

19. Jewell NP. Small-sample bias of point estimators of the odds ratio from matched sets. Biometrics. 1984;40(2):421-35.

20. Albert A, Anderson JA. On the existence of maximum likelihood estimates in logistic regression models. Biometrika. 1984;71(1):1-10.

21. Peduzzi P, Concato J, Kemper E, Holford TR, Feinstein AR. A simulation study of the number of events per variable in logistic regression analysis. J Clin Epidemiol. 1996;49(12):1373-9.

22. Vittinghoff E, McCulloch CE. Relaxing the rule of ten events per variable in logistic and Cox regression. Am J Epidemiol. 2007;165(6):710-8.

23. Steyerberg EW. Clinical Prediction Models: A Practical Approach to Development, Validation, and Updating: Springer International Publishing; 2019.

24. Akaike H. Information theory and an extension of the maximum likelihood principle. 2nd International Symposium on Information Theory. 1973:267-81.

25. Schwarz G. Estimating the dimension of a model. Ann Stat. 1978;6(2):461-4.

26. Rahman MS, Sultana M. Performance of Firth-and log F-type penalized methods in risk prediction for small or sparse binary data. BMC Med Res Methodol. 2017;17(1):33.

27. Steyerberg EW, Eijkemans MJC, Habbema JDF. Application of shrinkage techniques in logistic regression analysis: A case study. Stat Neerl. 2001;55(1):76-88.

28. R Core Team. R: a language and environment for statistical computing. R Foundation for Statistical Computing. 2018.

29. Heinze G, Ploner M. logistf: Firth's bias-reduced logistic regression. R package version 123. 2018.

30. Friedman J, Hastie T, Tibshirani R. Regularization paths for generalized linear models via coordinate descent. J Stat Softw. 2010;33(1):1-22.

31. Hastie T, Tibshirani R, Friedman JH. The Elements of Statistical Learning: Data Mining, Inference, and Prediction: Springer; 2009.

32. Mueller HS, Cohen LS, Braunwald E, Forman S, Feit F, Ross A, et al. Predictors of early morbidity and mortality after thrombolytic therapy of acute myocardial infarction. Analyses of patient subgroups in the Thrombolysis in Myocardial Infarction (TIMI) trial, phase II. Circulation. 1992;85(4):1254-64.

33. Dai B, Ding S, Wahba G. Multivariate Bernoulli distribution. Bernoulli. 2013;19(4):1465-83.

34. Barthélemy J, Suesse T. mipfp: an R package for multidimensional array fitting and simulating multivariate Bernoulli distributions. J Stat Softw. 2018;86(Code Snippet 2).

35. Tantithamthavorn C, McIntosh S, Hassan AE, Matsumoto K. An empirical comparison of model validation techniques for defect prediction models. IEEE Trans Softw Eng. 2017;43(1):1-18.

**Table 1.** Characteristics of the GUSTO-I West region dataset.

| | |
|---|---|
| N | 2,188 |
| Outcome | |
|   30-day mortality | 6.2% |
| Covariates | |
|   Age >65 years | 38.4% |
|   Female gender | 24.9% |
|   Diabetes | 14.3% |
|   Hypotension (systolic blood pressure <100 mmHg) | 9.6% |
|   Tachycardia (pulse >80 bpm) | 33.4% |
|   High risk (anterior infarct location/previous MI) | 48.7% |
|   Shock (Killip class III/IV) | 1.5% |
|   Time to relief of chest pain >1 hour | 60.9% |
|   Previous MI | 17.1% |
|   Height (cm) | 172.1 ± 10.1 |
|   Weight (kg) | 82.9 ± 17.7 |
|   Hypertension history | 40.4% |
|   Ex-smoker | 30.8% |
|   Current smoker | 27.9% |
|   Hypercholesterolemia | 40.5% |
|   Previous angina pectoris | 34.1% |
|   Family history of MI | 47.6% |
|   ST elevation in >4 leads | 35.6% |

**Table 2.** Regression coefficients estimates and predictive performance measures for the GUSTO-I trial West region dataset.

| | 8-Predictor Model | | | | | | | 17-Predictor Model | | | | | | |
|---|---|---|---|---|---|---|---|---|---|---|---|---|---|---|
| | ML | Firth | Ridge | Lasso | Elastic-net | Backward (AIC) | Backward (P < 0.05) | ML | Firth | Ridge | Lasso | Elastic-net | Backward (AIC) | Backward (P < 0.05) |
| Coefficients: | | | | | | | | | | | | | | |
| Intercept | −5.092 | −5.034 | −4.787 | −4.933 | −4.886 | −4.927 | −4.927 | −5.090 | −4.983 | −3.434 | −3.494 | −3.486 | −3.494 | −2.853 |
| Age >65 years | 1.637 | 1.616 | 1.424 | 1.578 | 1.535 | 1.631 | 1.631 | 1.429 | 1.399 | 1.161 | 1.336 | 1.324 | 1.495 | 1.532 |
| Female gender | 0.622 | 0.620 | 0.592 | 0.586 | 0.586 | 0.624 | 0.624 | 0.490 | 0.487 | 0.380 | 0.288 | 0.289 | 0.368 | . |
| Diabetes | 0.069 | 0.083 | 0.078 | 0.024 | 0.035 | . | . | 0.153 | 0.164 | 0.131 | . | . | . | . |
| Hypotension | 1.218 | 1.215 | 1.102 | 1.164 | 1.145 | 1.252 | 1.252 | 1.192 | 1.178 | 1.037 | 1.036 | 1.030 | 1.230 | 1.227 |
| Tachycardia | 0.650 | 0.645 | 0.574 | 0.608 | 0.597 | 0.661 | 0.661 | 0.653 | 0.643 | 0.533 | 0.530 | 0.526 | 0.669 | 0.717 |
| High risk | 0.847 | 0.835 | 0.748 | 0.796 | 0.781 | 0.855 | 0.855 | 0.403 | 0.397 | 0.390 | 0.372 | 0.372 | 0.414 | 2.748 |
| Shock | 2.395 | 2.362 | 2.339 | 2.362 | 2.354 | 2.424 | 2.424 | 2.685 | 2.608 | 2.508 | 2.460 | 2.455 | 2.662 | 0.779 |
| No relief of chest pain | 0.263 | 0.255 | 0.237 | 0.219 | 0.221 | . | . | 0.233 | 0.223 | 0.200 | 0.107 | 0.107 | . | . |
| Previous MI | | | | | | | | 0.505 | 0.495 | 0.437 | 0.378 | 0.376 | 0.586 | . |
| Height | | | | | | | | 0.008 | 0.008 | −0.001 | . | . | . | . |
| Weight | | | | | | | | −0.019 | −0.018 | −0.015 | −0.014 | −0.014 | −0.018 | −0.023 |
| Hypertension history | | | | | | | | −0.165 | −0.159 | −0.123 | . | . | . | . |
| Ex-smoker | | | | | | | | 0.174 | 0.169 | 0.147 | . | . | . | . |
| Current smoker | | | | | | | | 0.247 | 0.241 | 0.231 | 0.057 | 0.059 | . | . |
| Hypercholesterolemia | | | | | | | | −0.064 | −0.060 | −0.064 | . | . | . | . |
| Previous angina pectoris | | | | | | | | 0.246 | 0.243 | 0.235 | 0.151 | 0.151 | . | . |
| Family history of MI | | | | | | | | −0.015 | −0.014 | −0.036 | . | . | . | . |
| ST elevation in >4 leads | | | | | | | | 0.583 | 0.571 | 0.479 | 0.434 | 0.430 | 0.601 | 0.752 |
| $C$-statistics: | | | | | | | | | | | | | | |
| Apparent | 0.819 | 0.819 | 0.819 | 0.819 | 0.819 | 0.820 | 0.820 | 0.832 | 0.832 | 0.831 | 0.831 | 0.831 | 0.829 | 0.824 |
| Harrell | 0.810 | 0.810 | 0.812 | 0.810 | 0.810 | 0.811 | 0.810 | 0.811 | 0.811 | 0.812 | 0.812 | 0.812 | 0.810 | 0.806 |
| .632 | 0.811 | 0.811 | 0.812 | 0.811 | 0.810 | 0.811 | 0.809 | 0.811 | 0.811 | 0.813 | 0.813 | 0.813 | 0.809 | 0.808 |
| .632+ | 0.810 | 0.811 | 0.812 | 0.811 | 0.810 | 0.811 | 0.809 | 0.810 | 0.810 | 0.812 | 0.812 | 0.812 | 0.809 | 0.808 |

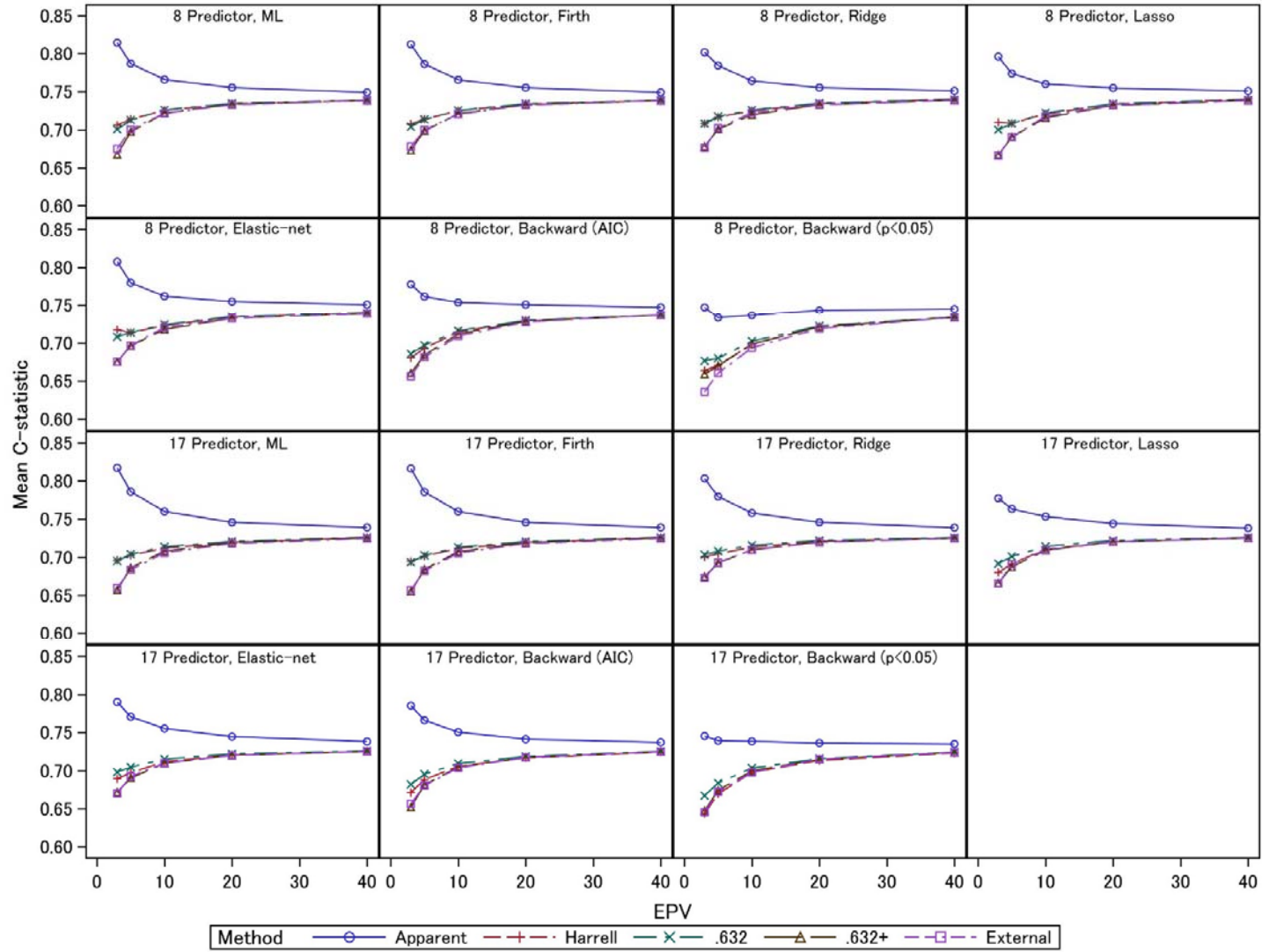

**Figure 1.** Simulation results: apparent, external, and optimism-corrected *C*-statistics (scenario 2 and event fraction = 0.5).

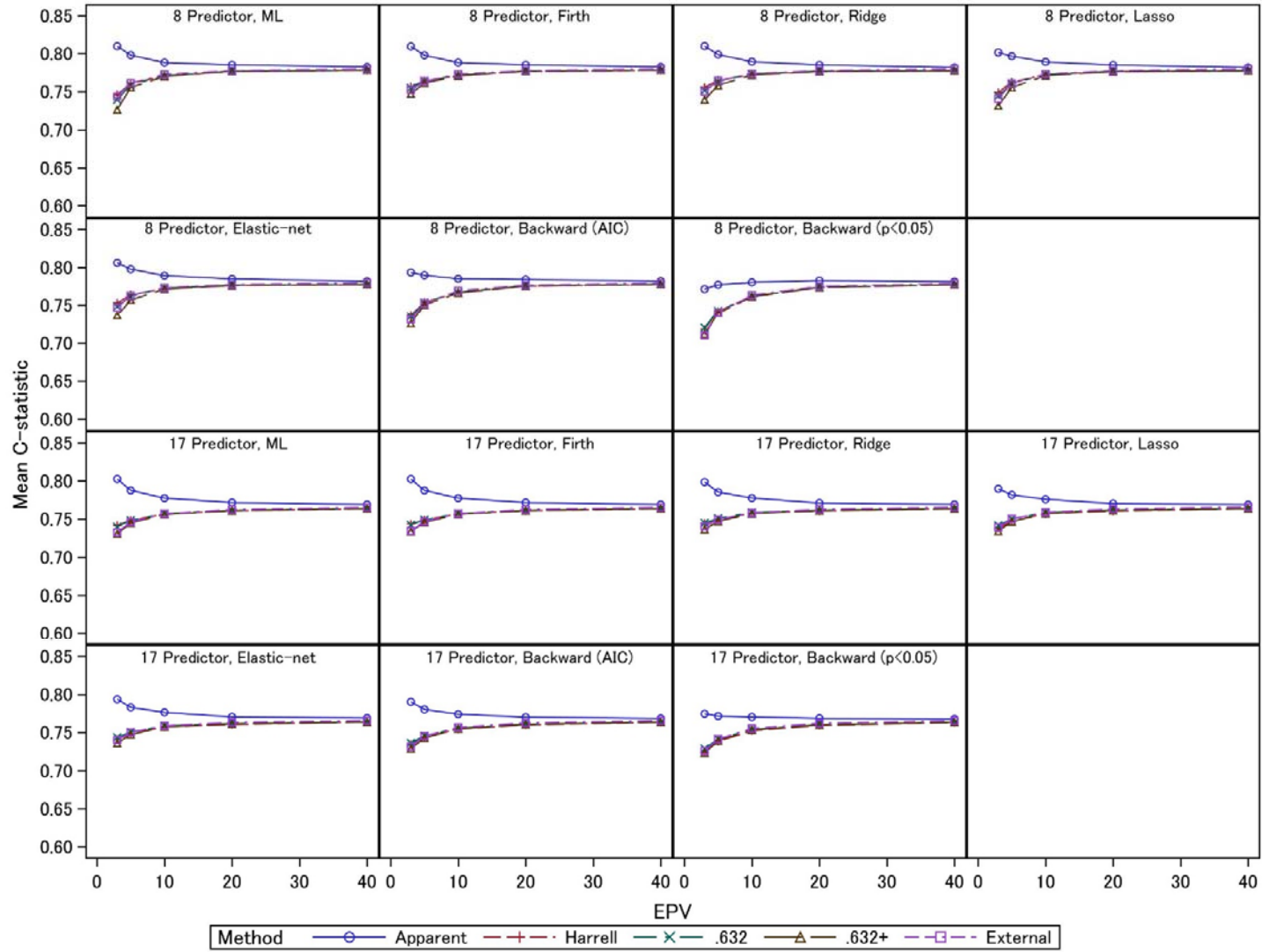

**Figure 2**. Simulation results: apparent, external, and optimism-corrected *C*-statistics (scenario 2 and event fraction = 0.0625).

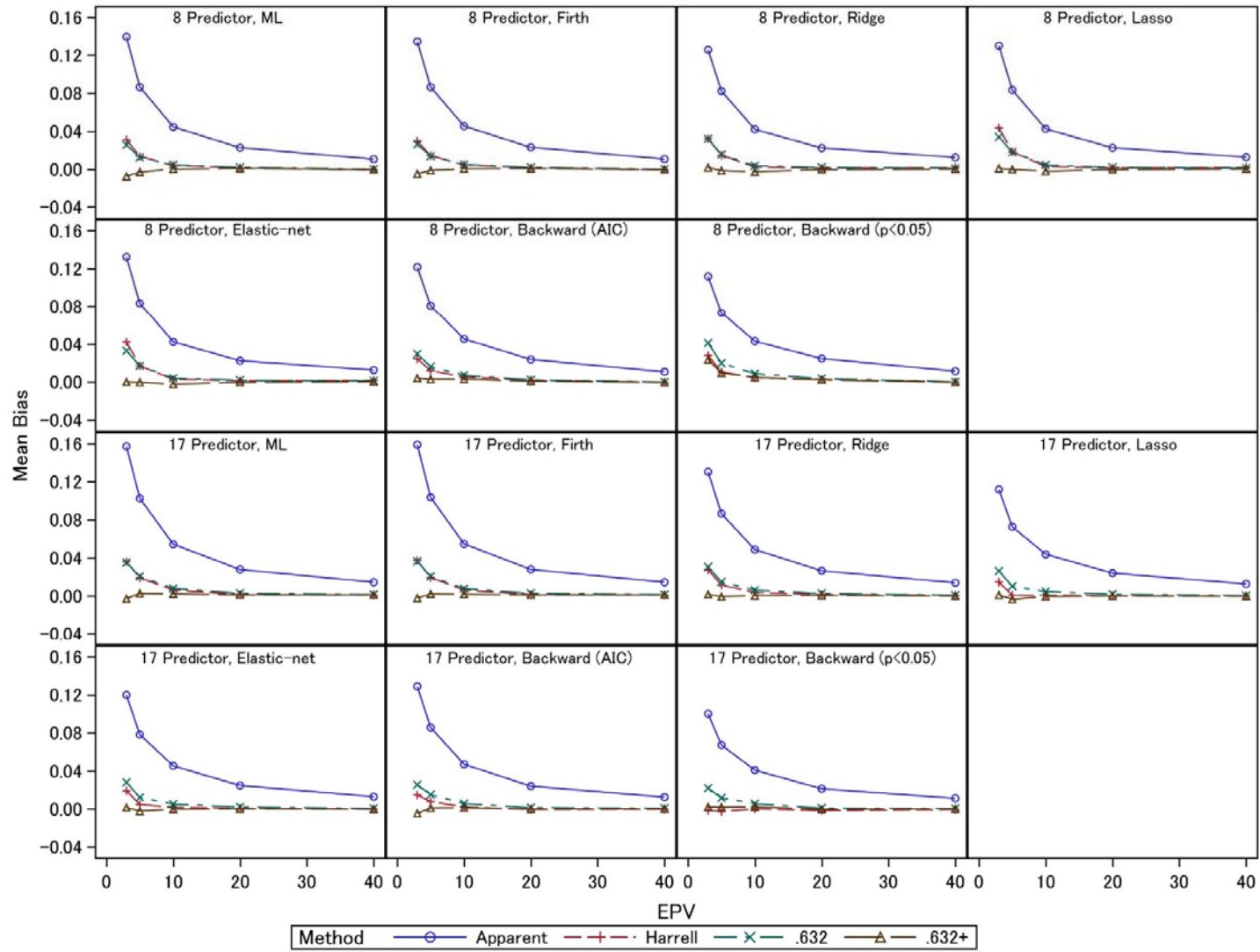

**Figure 3.** Simulation results: bias in the apparent and optimism-corrected *C*-statistics (scenario 2 and event fraction = 0.5).

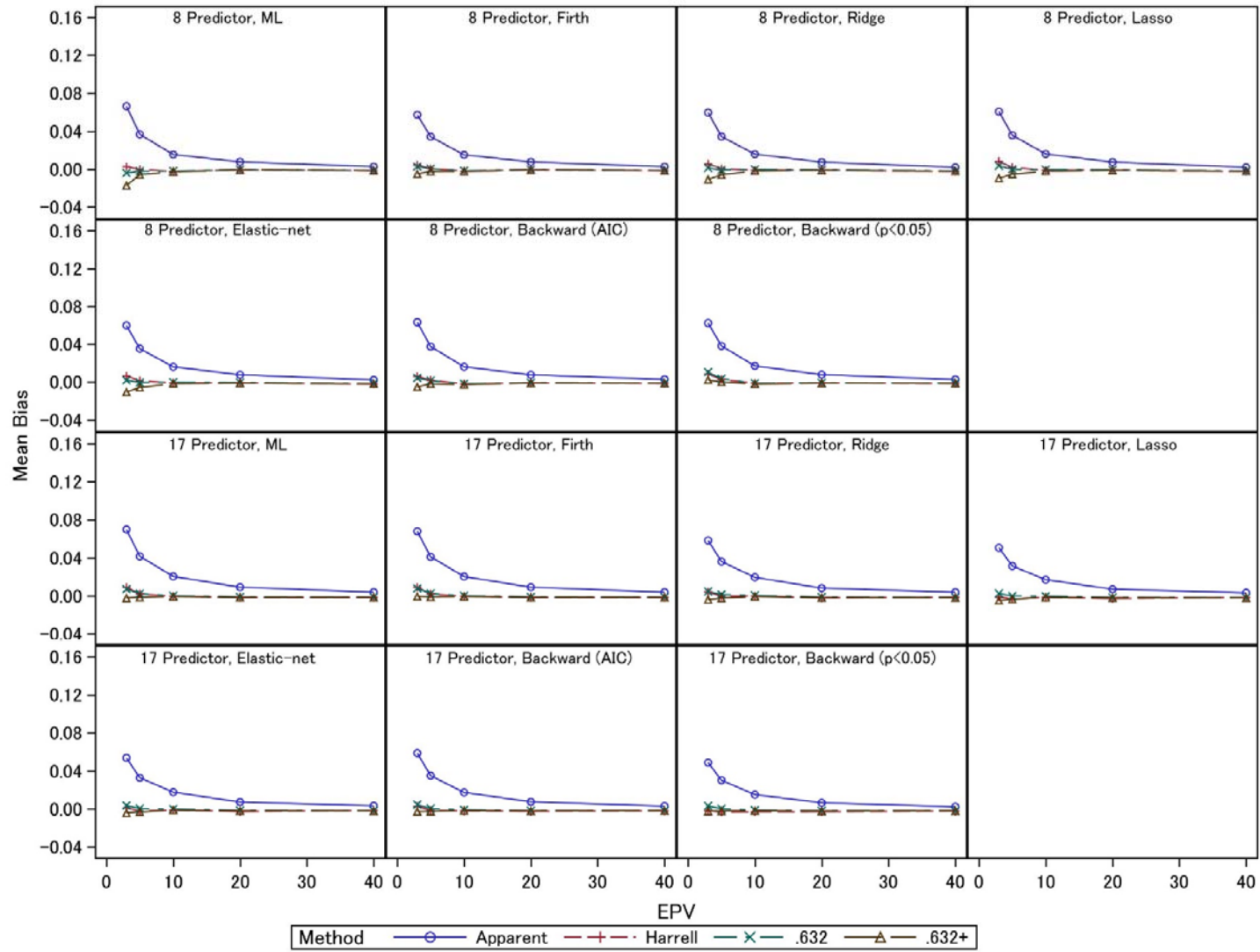

**Figure 4.** Simulation results: bias in the apparent and optimism-corrected *C*-statistics (scenario 2 and event fraction = 0.0625).

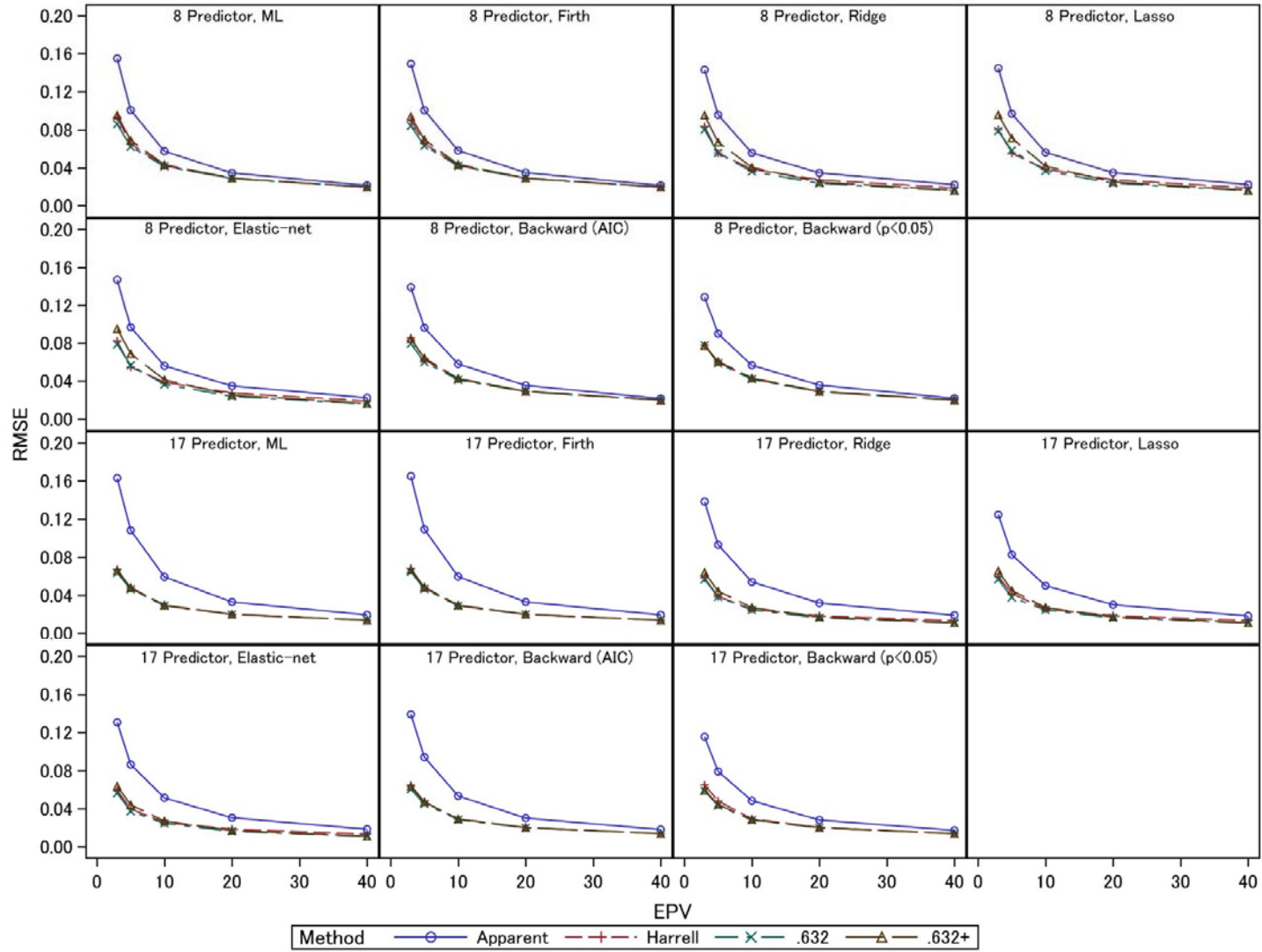

**Figure 5**. Simulation results: RMSE in the apparent and optimism-corrected *C*-statistics (scenario 2 and event fraction = 0.5).

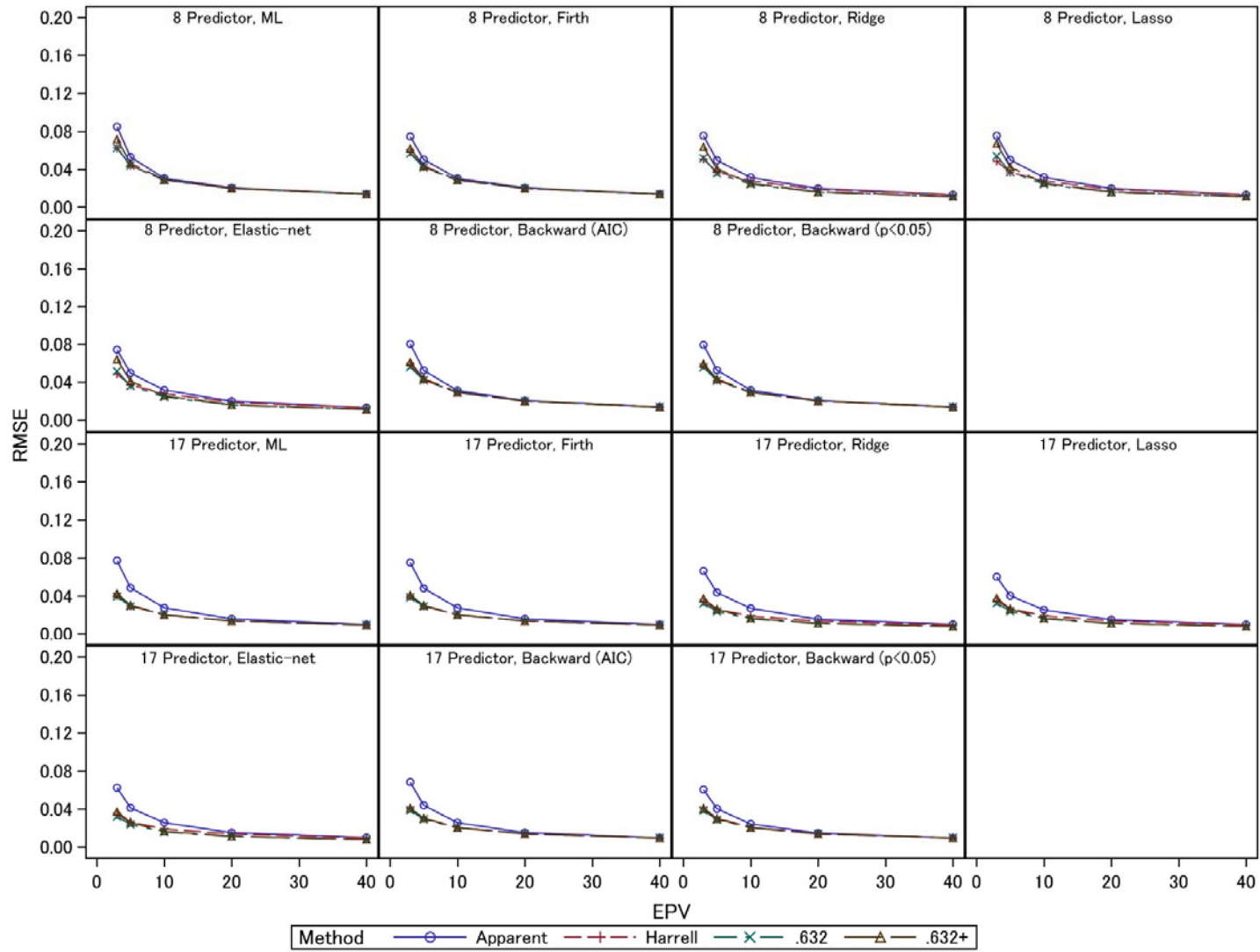

**Figure 6.** Simulation results: RMSE in the apparent and optimism-corrected *C*-statistics (scenario 2 and event fraction = 0.0625).

# Re-evaluation of the comparative effectiveness of bootstrap-based optimism correction methods in the development of multivariable clinical prediction models

Katsuhiro Iba[1,2], Tomohiro Shinozaki[3], Kazushi Maruo[4] and Hisashi Noma[5]

[1] *Department of Statistical Science, School of Multidisciplinary Sciences, The Graduate University for Advanced Studies, Tokyo, Japan*
[2] *Office of Biostatistics, Department of Biometrics, Headquarters of Clinical Development, Otsuka Pharmaceutical Co., Ltd., Tokyo, Japan*
[3] *Department of Information and Computer Technology, Faculty of Engineering, Tokyo University of Science, Tokyo, Japan*
[4] *Department of Biostatistics, Faculty of Medicine, University of Tsukuba, Ibaraki, Japan.*
[5] *Department of Data Science, The Institute of Statistical Mathematics, Tokyo, Japan*

## e-Appendix A: All simulation results

We present the simulation results for all scenarios, methods, and modelling strategies. The results are provided as follows: ML (e-Figures 1 to 3), Firth (e-Figures 4 to 6), ridge (e-Figures 7 to 9), lasso (e-Figures 10 to 12), elastic-net (e-Figures 13 to 15), backward stepwise selections based on AIC (e-Figures 16 to 18) and based on significance level at 0.05 (e-Figures 19 to 21).

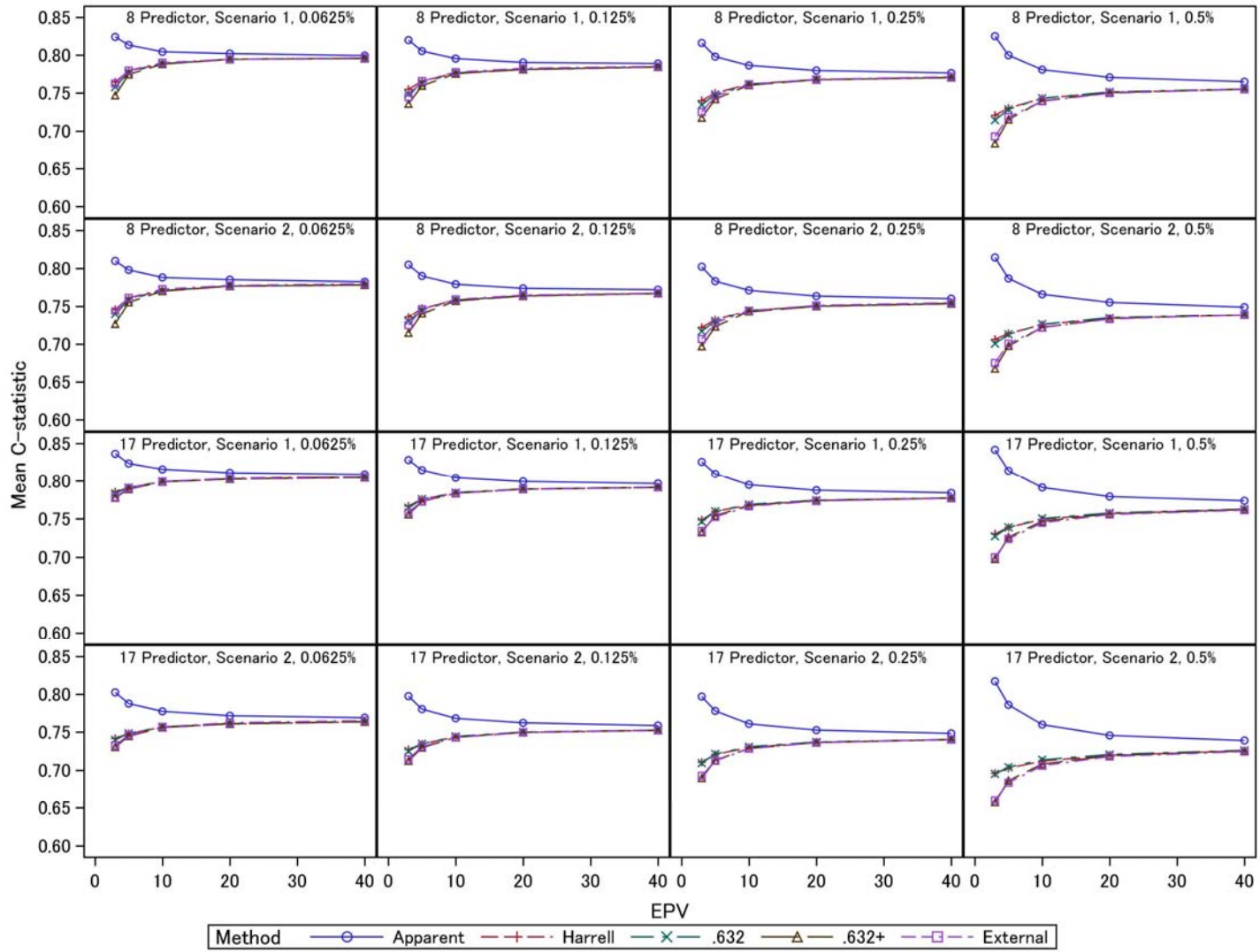

**e-Figure 1.** Simulation results: apparent, external, and optimism corrected *C*-statistics (ML estimation)

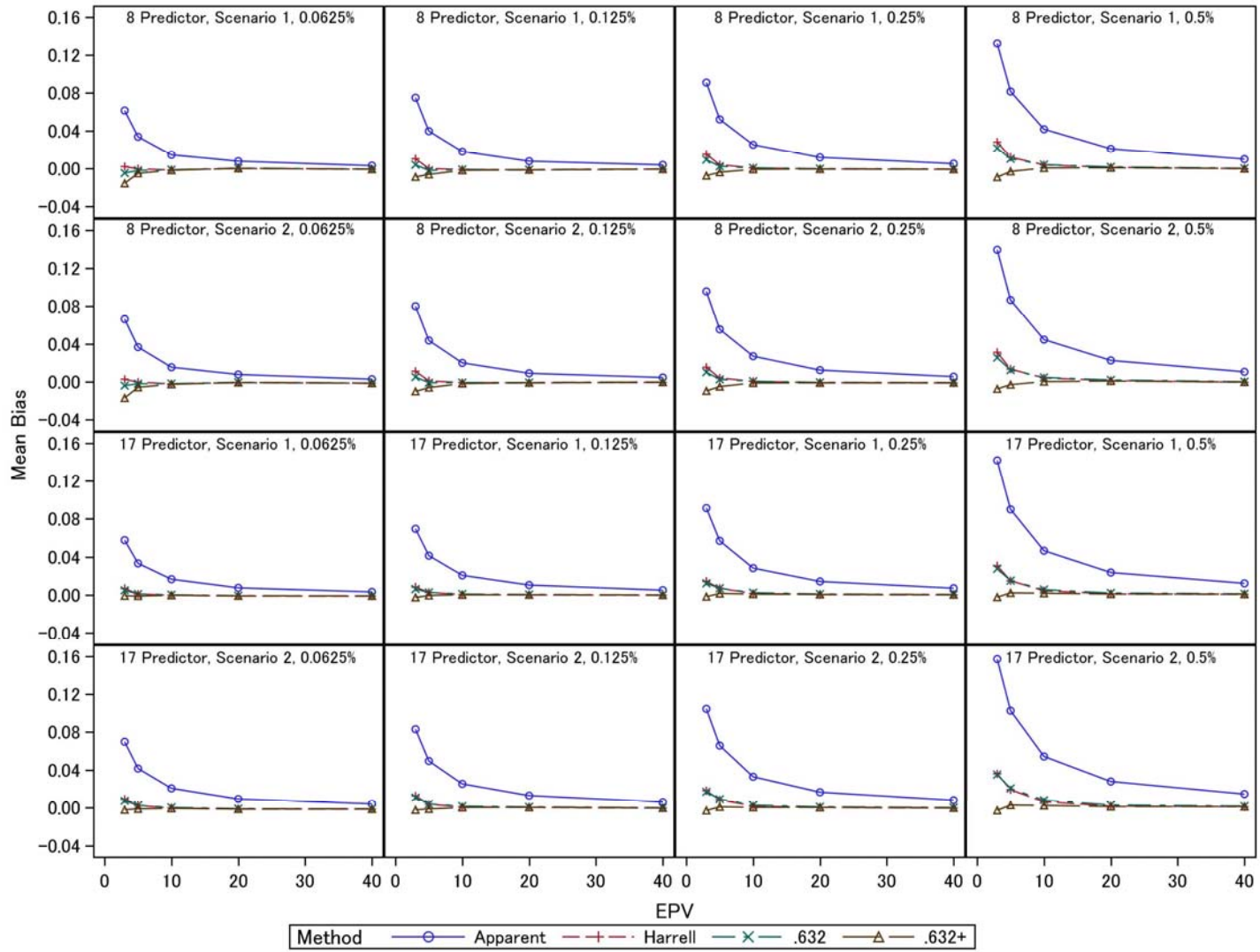

**e-Figure 2.** Simulation results: bias in apparent and optimism corrected *C*-statistics (ML estimation)

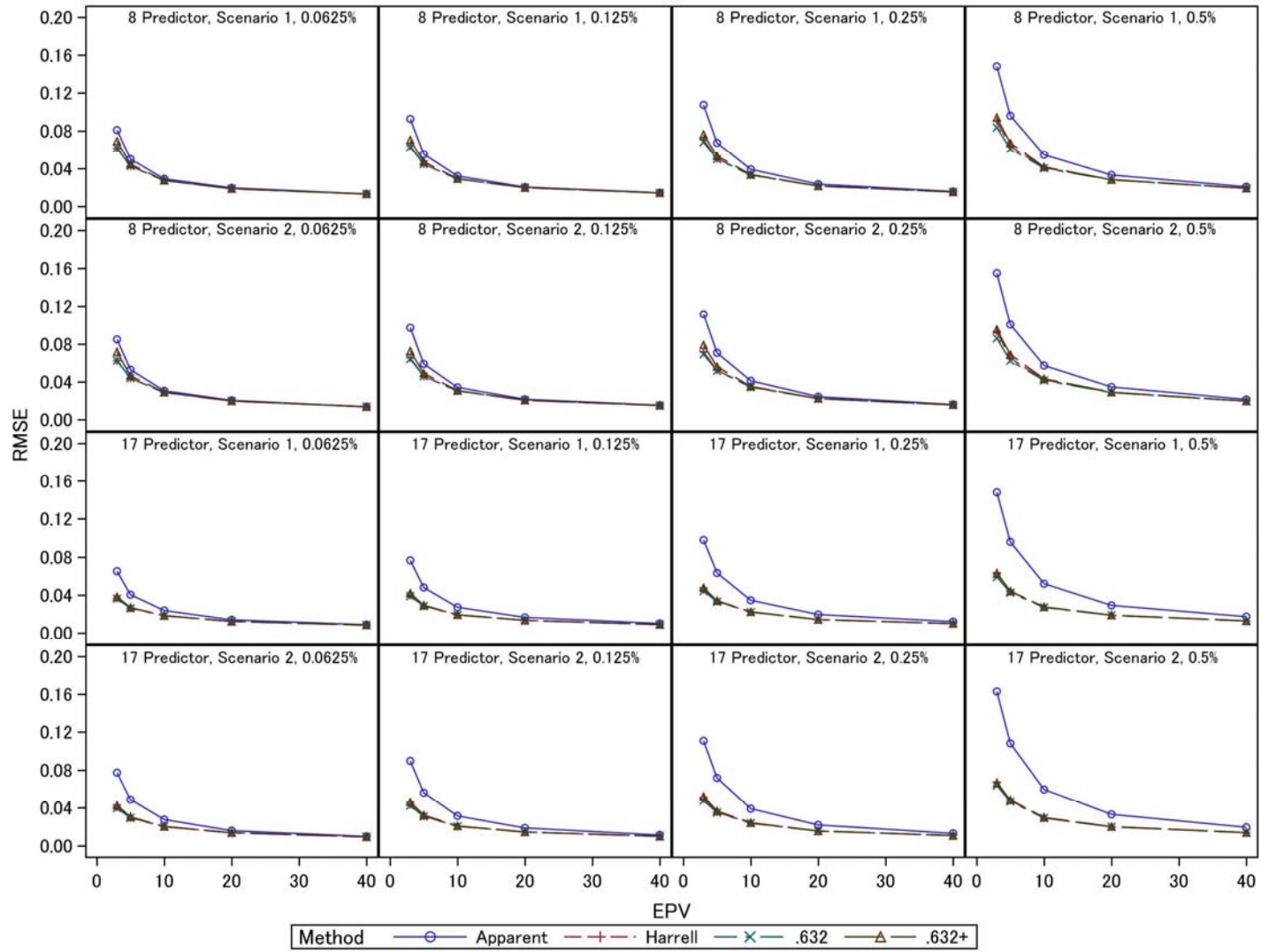

**e-Figure 3.** Simulation results: RMSE in apparent and optimism corrected *C*-statistics (ML estimation)

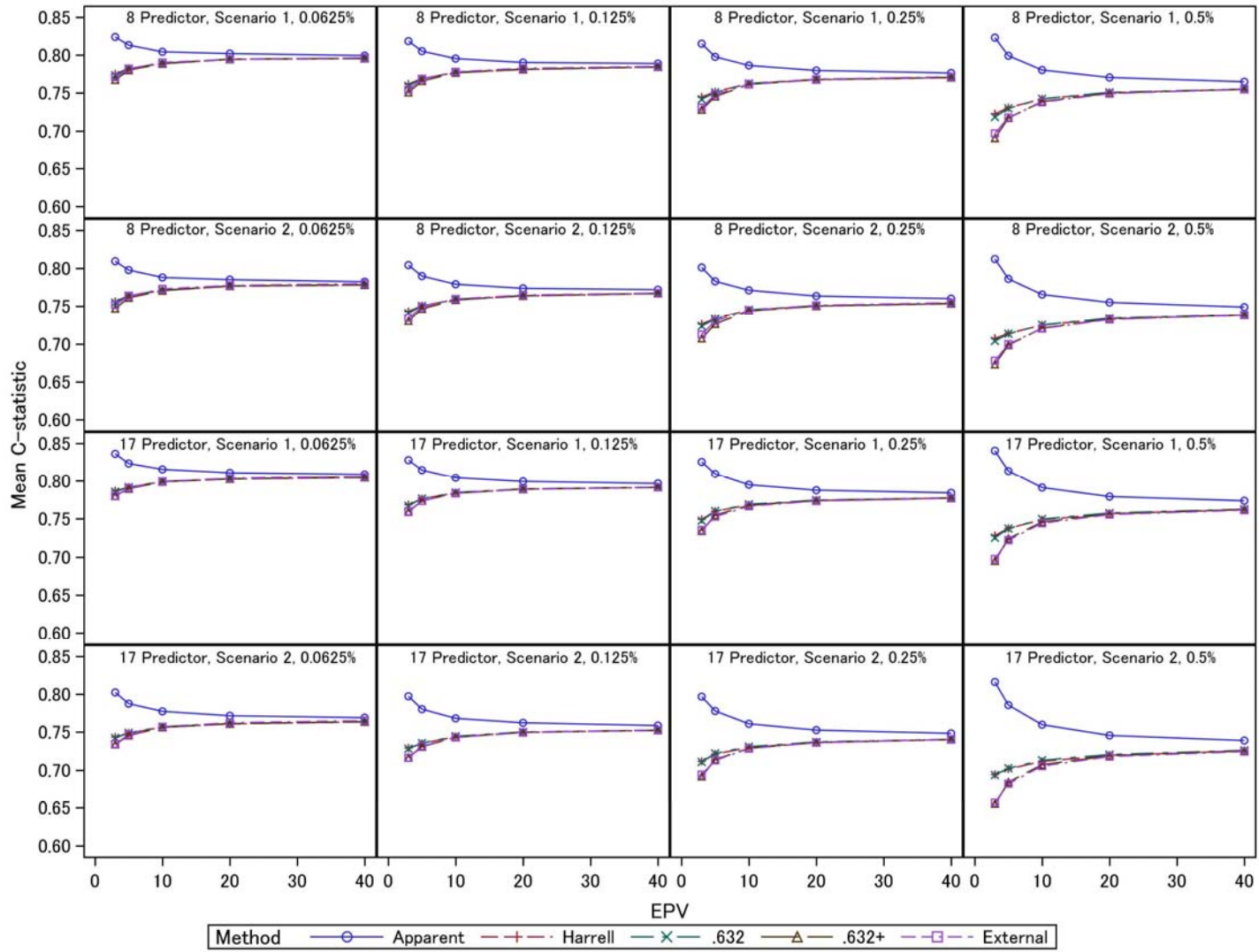

**e-Figure 4.** Simulation results: apparent, external, and optimism corrected *C*-statistics (Firth's penalized likelihood method)

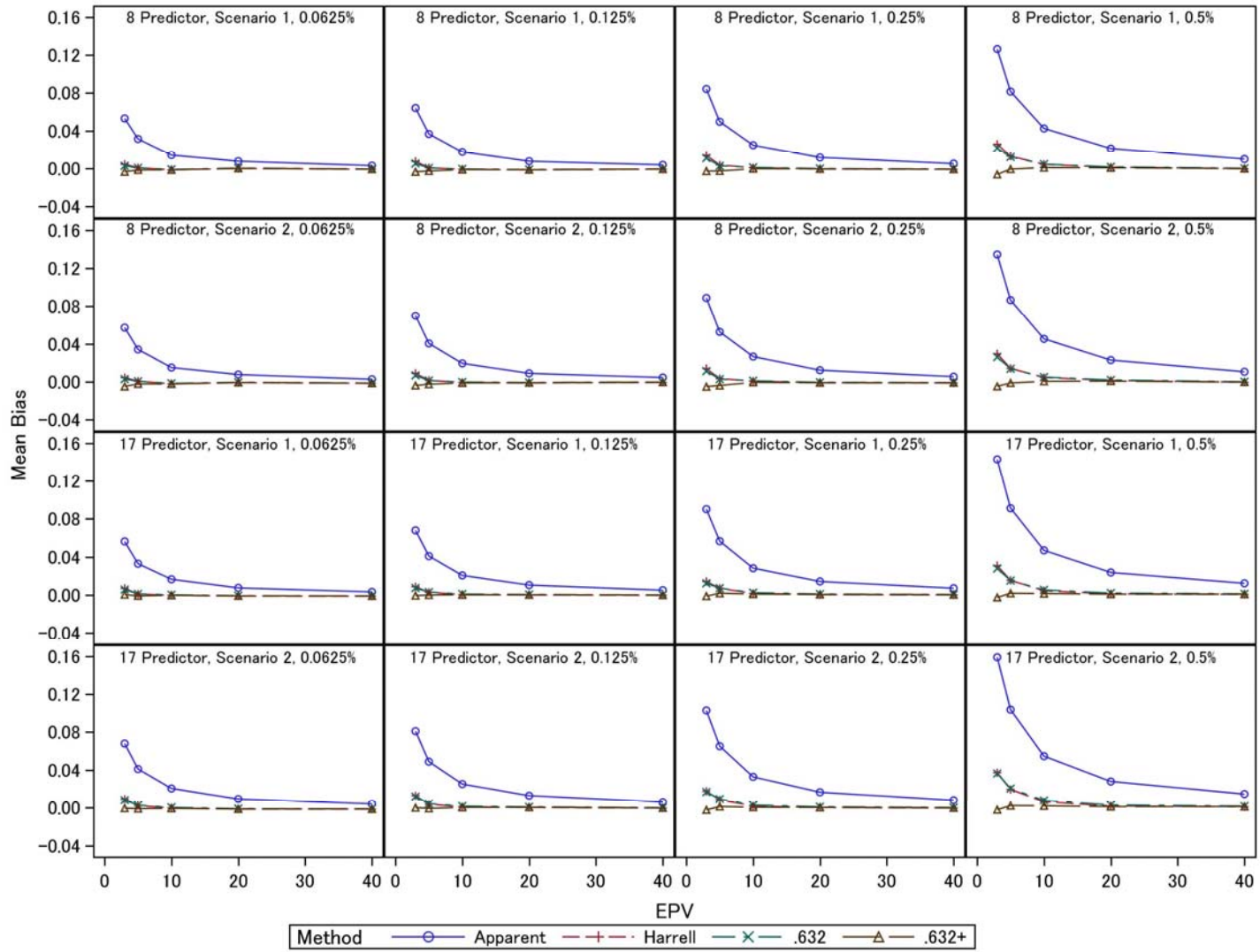

**e-Figure 5.** Simulation results: bias in apparent and optimism corrected *C*-statistics (Firth's penalized likelihood method)

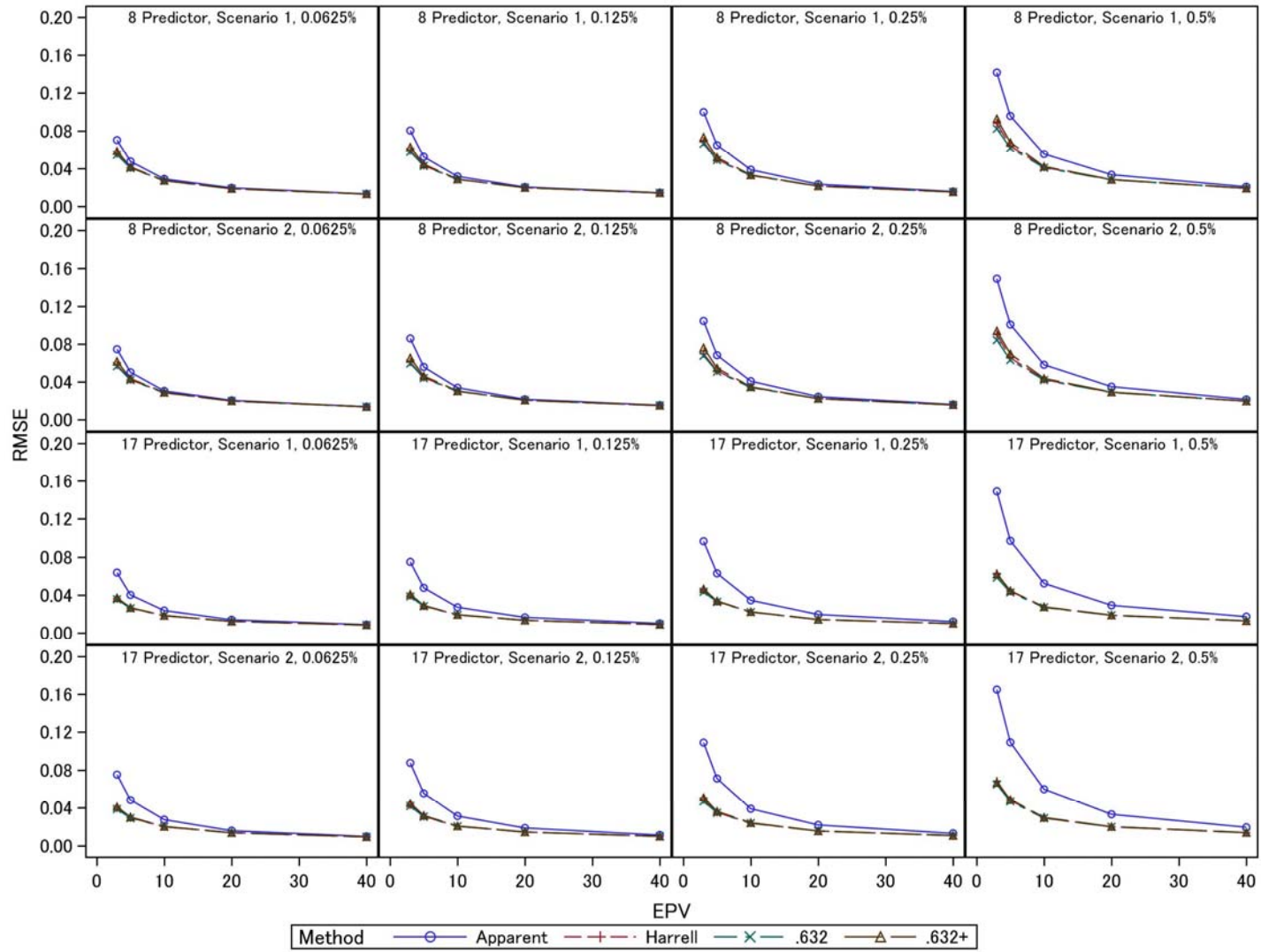

**e-Figure 6.** Simulation results: RMSE in apparent and optimism corrected *C*-statistics (Firth's penalized likelihood method)

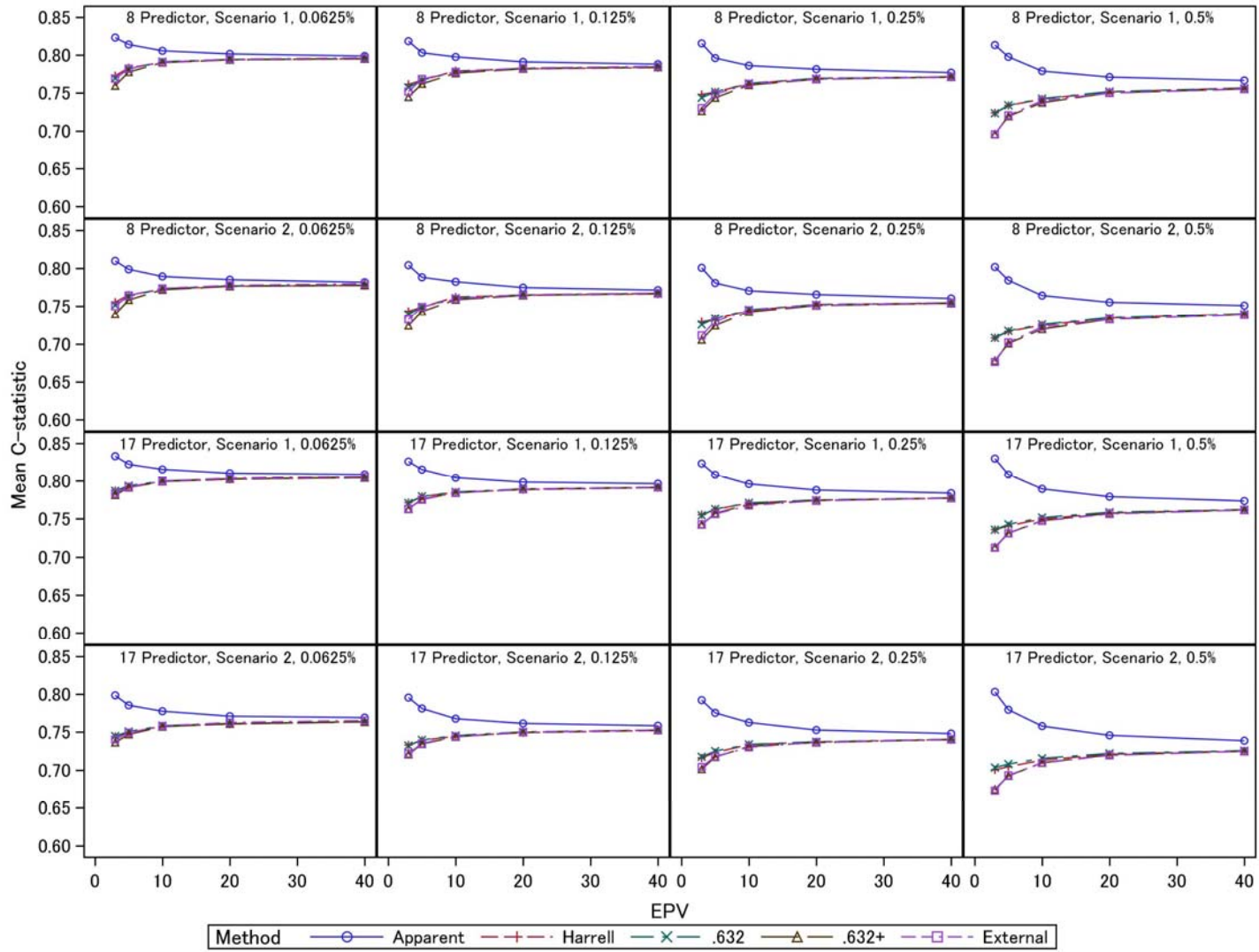

**e-Figure 7.** Simulation results: apparent, external, and optimism corrected *C*-statistics (ridge estimation)

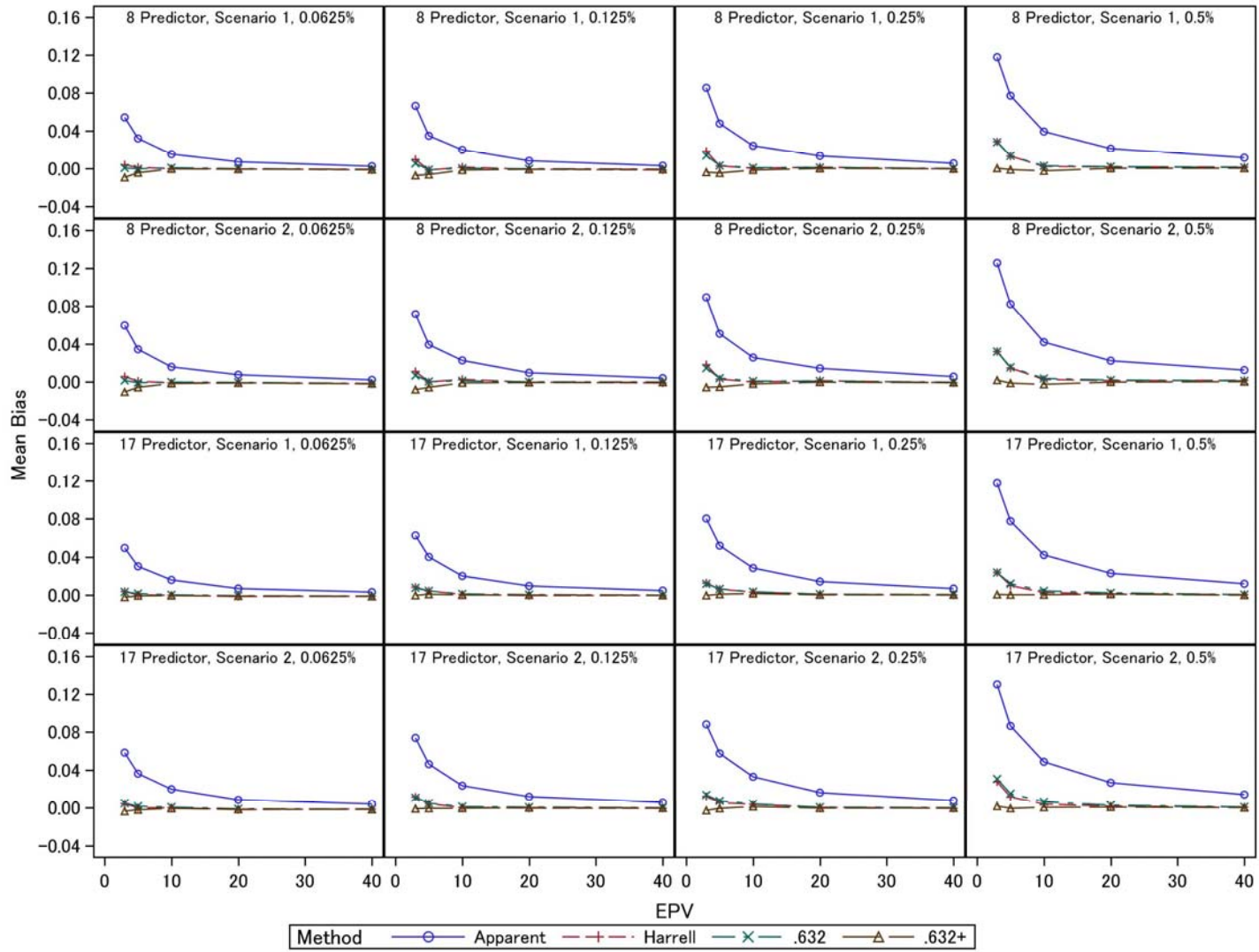

**e-Figure 8.** Simulation results: bias in apparent and optimism corrected *C*-statistics (ridge estimation)

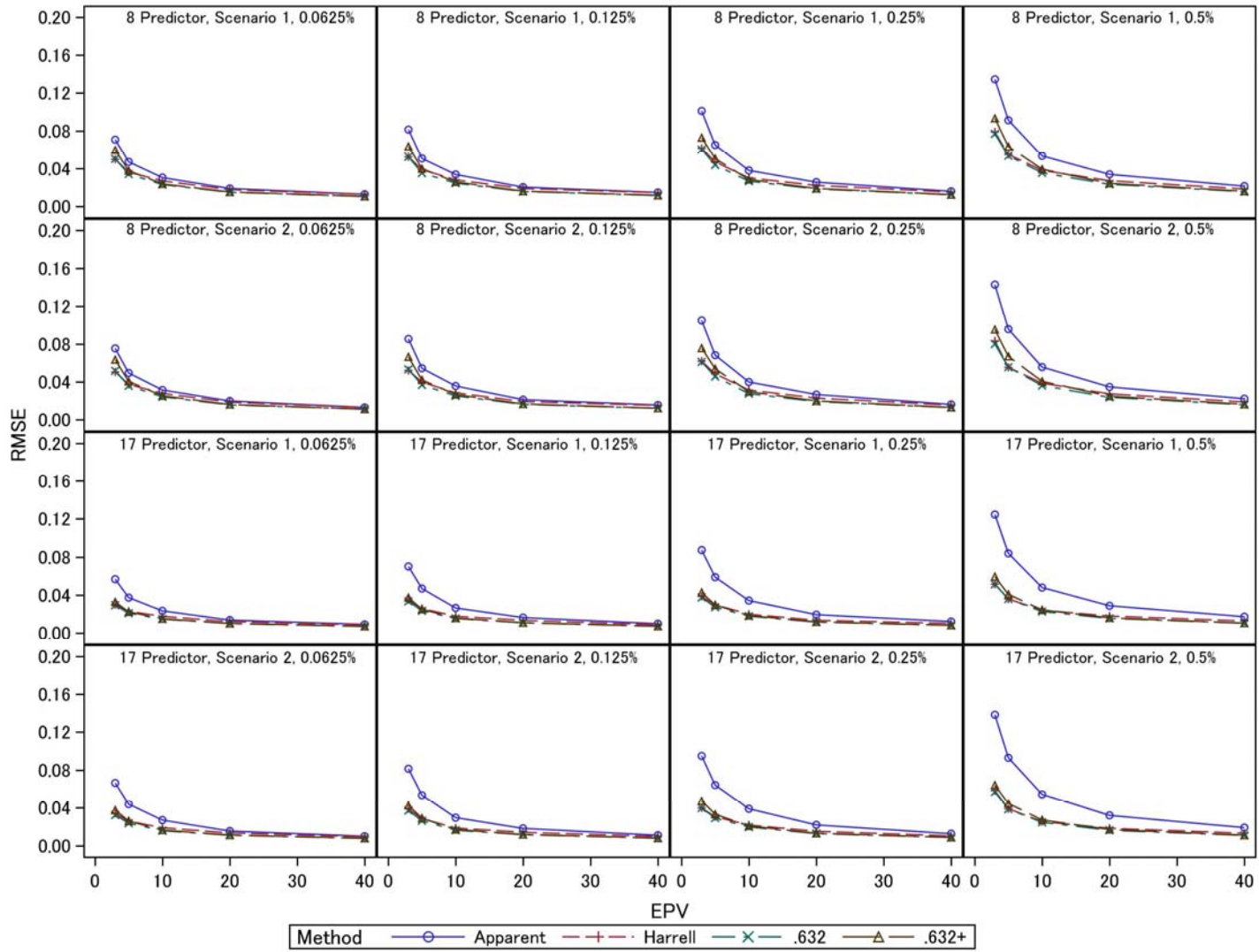

**e-Figure 9.** Simulation results: RMSE in apparent and optimism corrected *C*-statistics (ridge estimation)

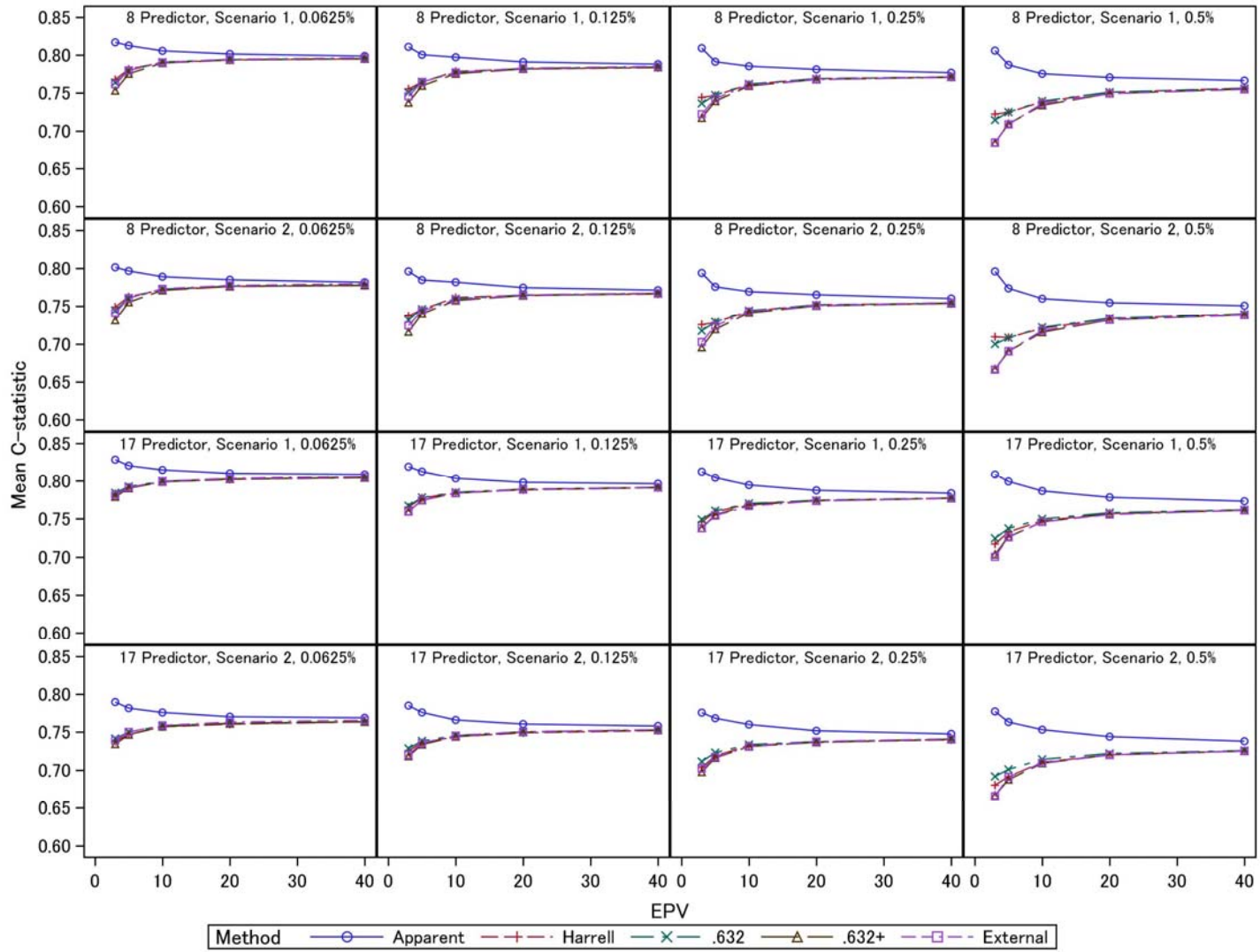

**e-Figure 10.** Simulation results: apparent, external, and optimism corrected *C*-statistics (lasso estimation)

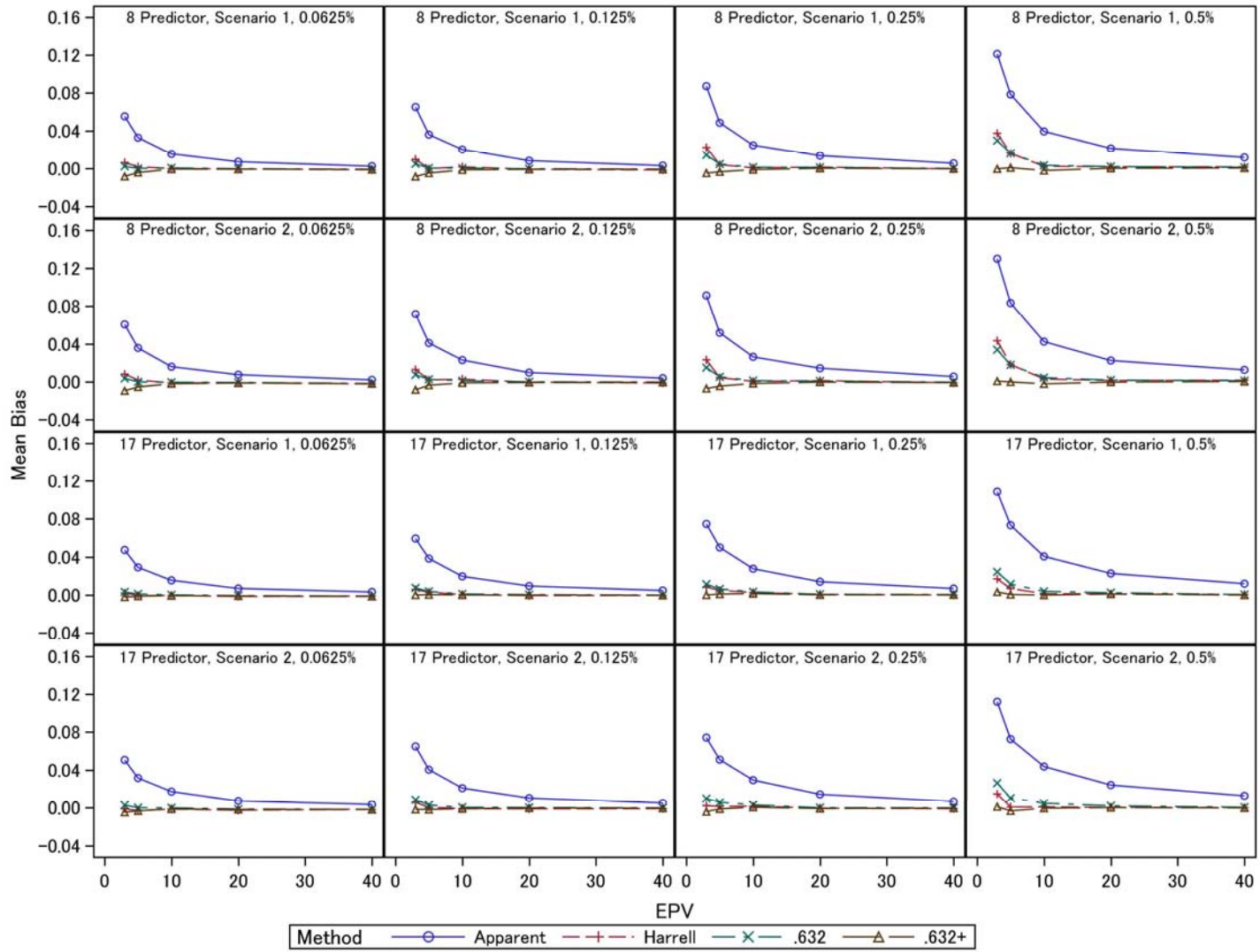

**e-Figure 11.** Simulation results: bias in apparent and optimism corrected *C*-statistics (lasso estimation)

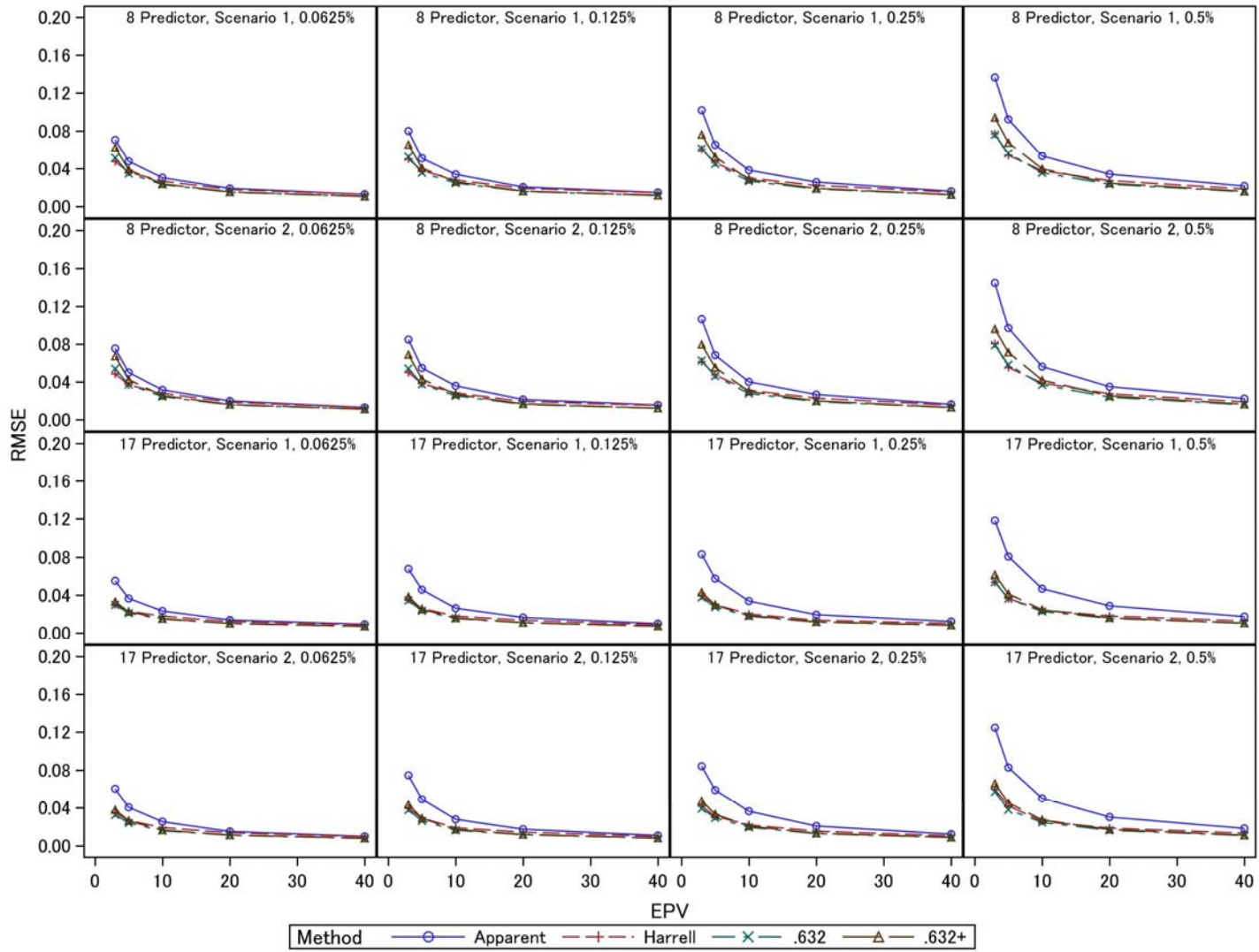

**e-Figure 12.** Simulation results: RMSE in apparent and optimism corrected *C*-statistics (lasso estimation)

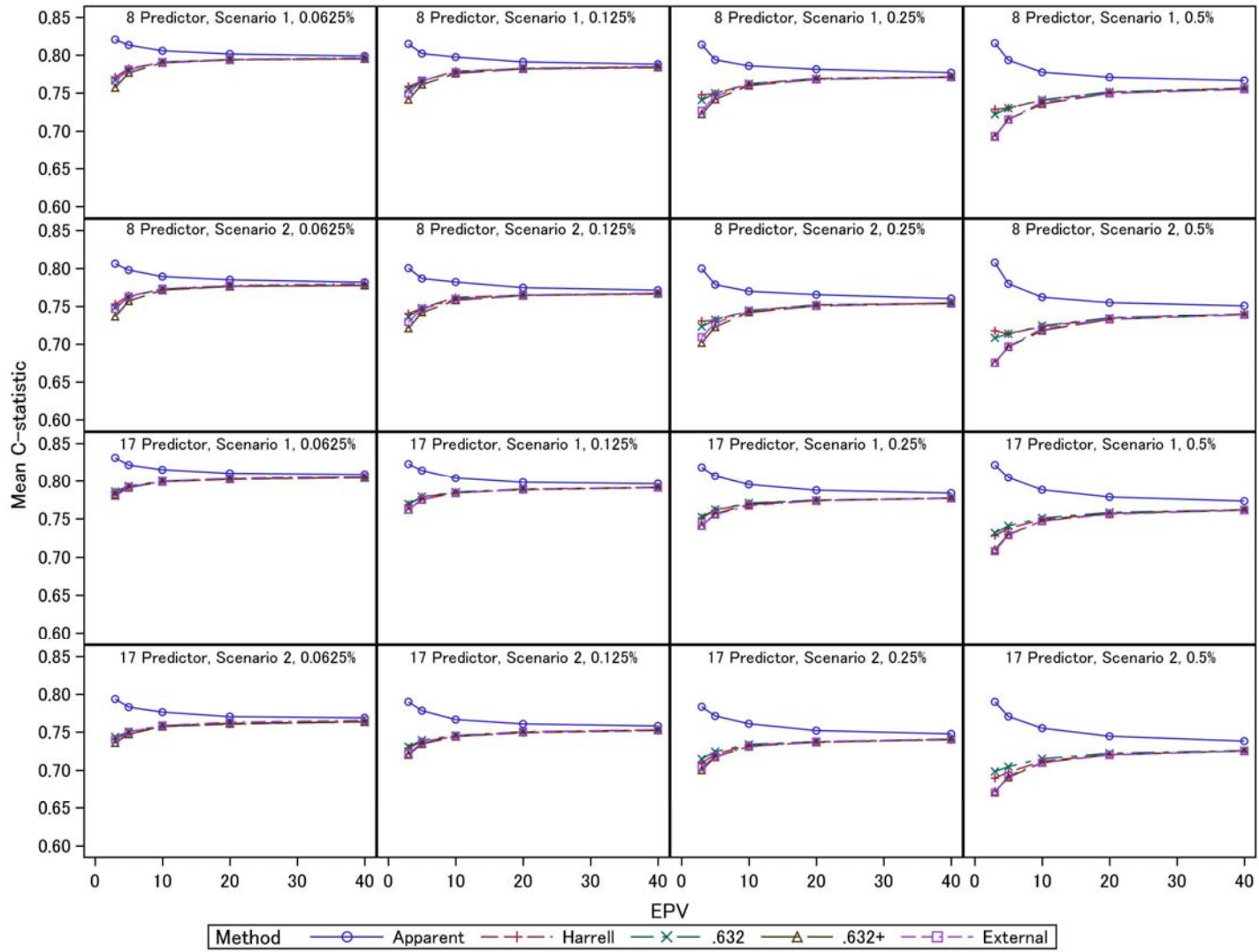

**e-Figure 13.** Simulation results: apparent, external, and optimism corrected *C*-statistics (elastic-net estimation)

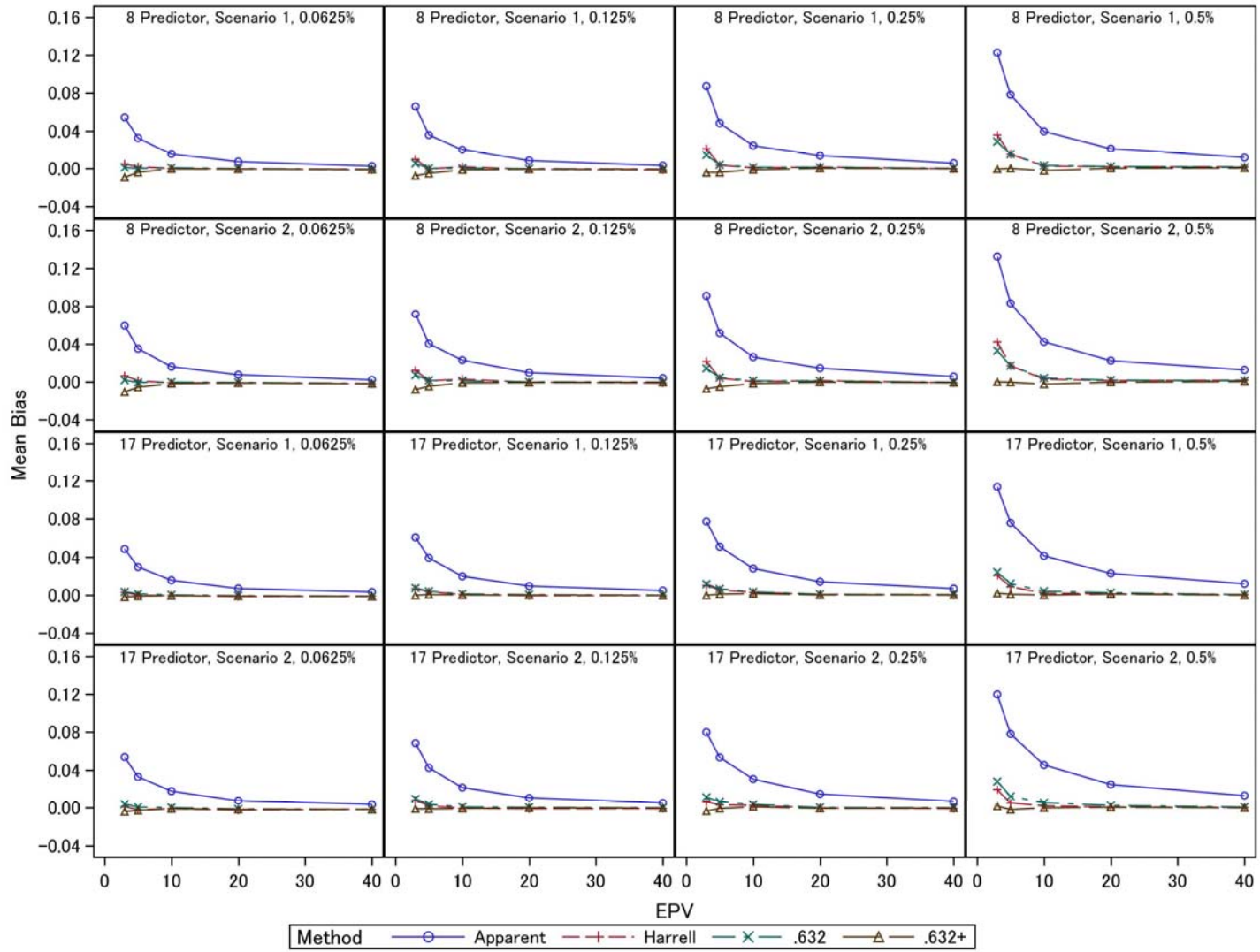

**e-Figure 14.** Simulation results: bias in apparent and optimism corrected *C*-statistics (elastic-net estimation)

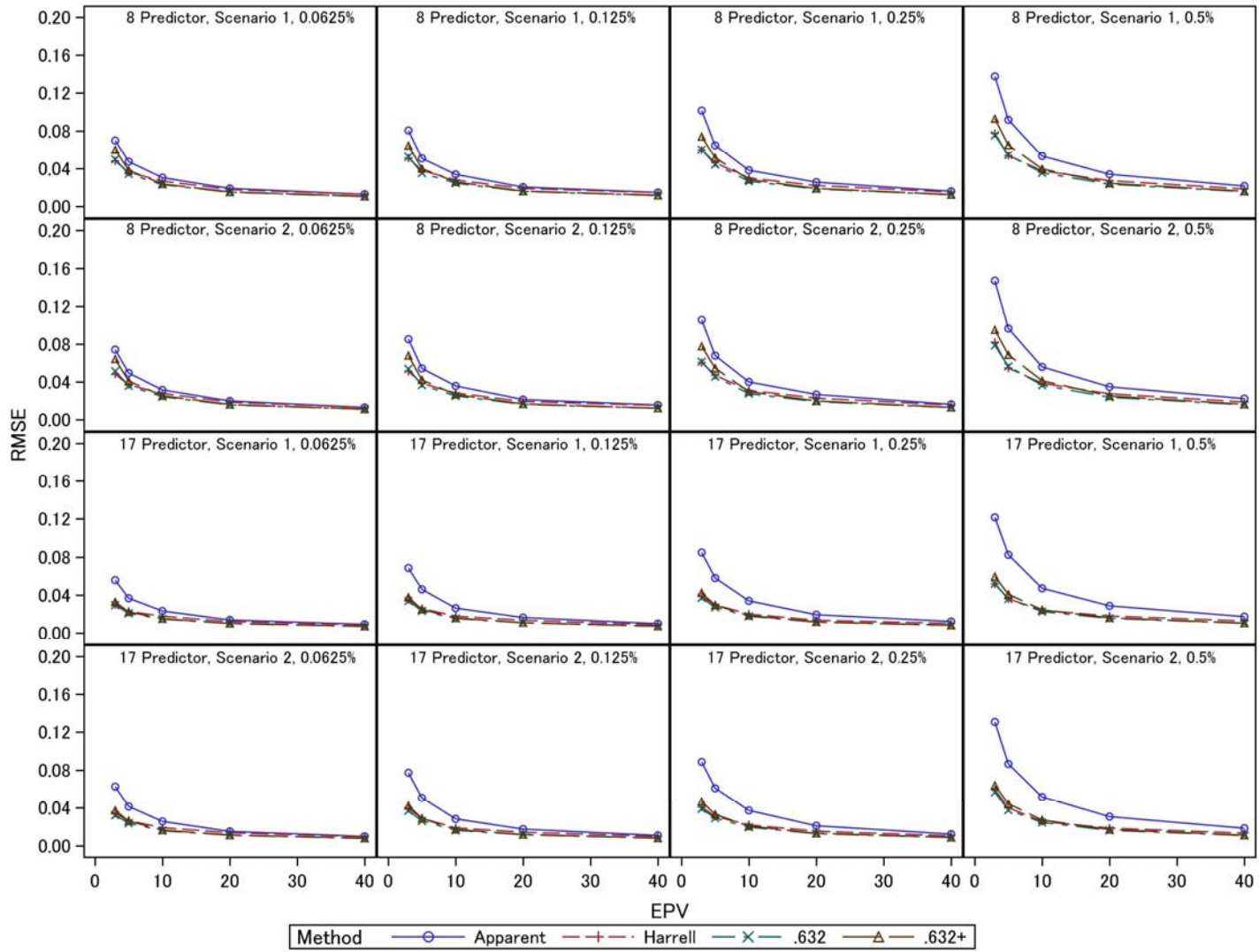

**e-Figure 15.** Simulation results: RMSE in apparent and optimism corrected *C*-statistics (elastic-net estimation)

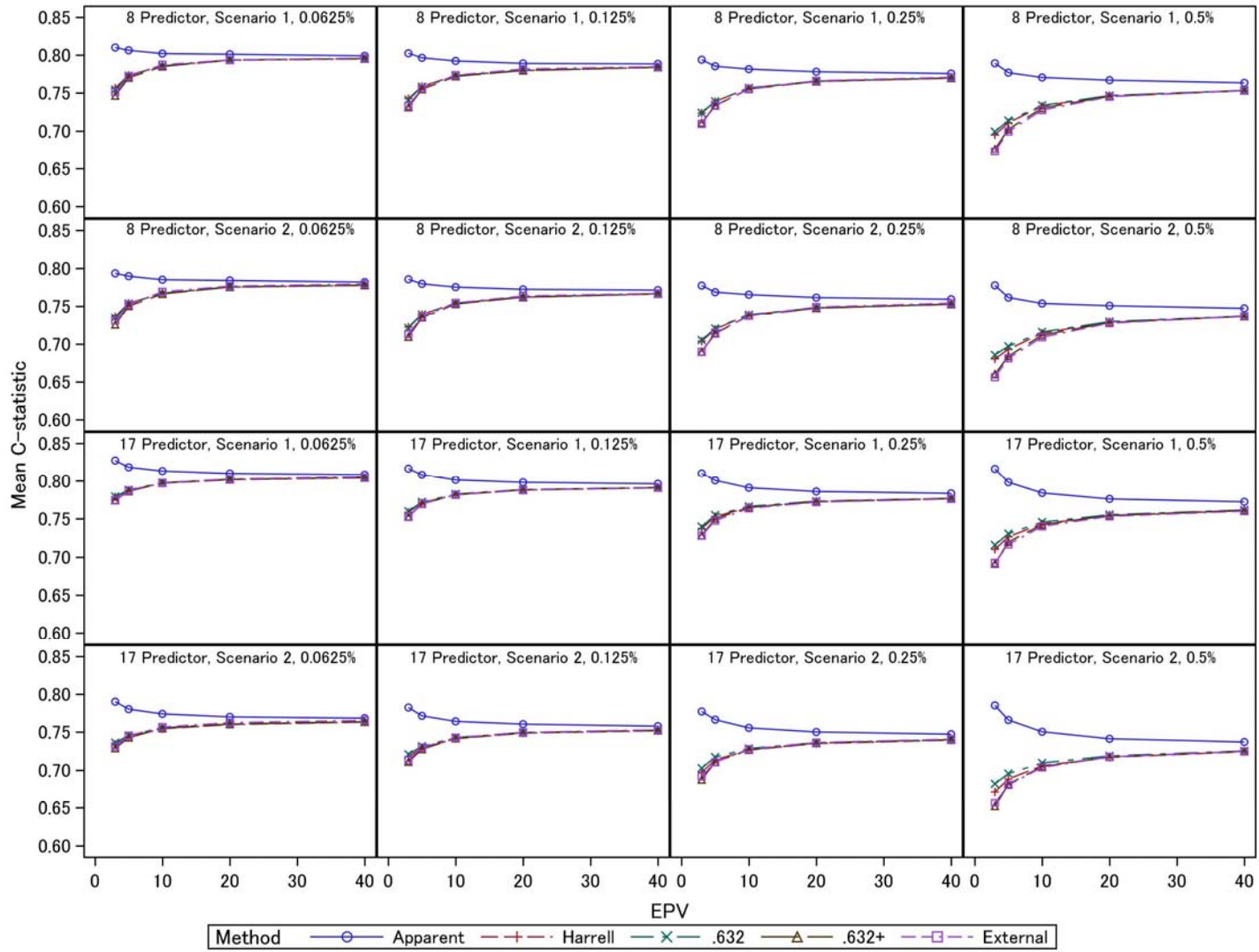

**e-Figure 16.** Simulation results: apparent, external, and optimism corrected *C*-statistics (Stepwise selection; AIC)

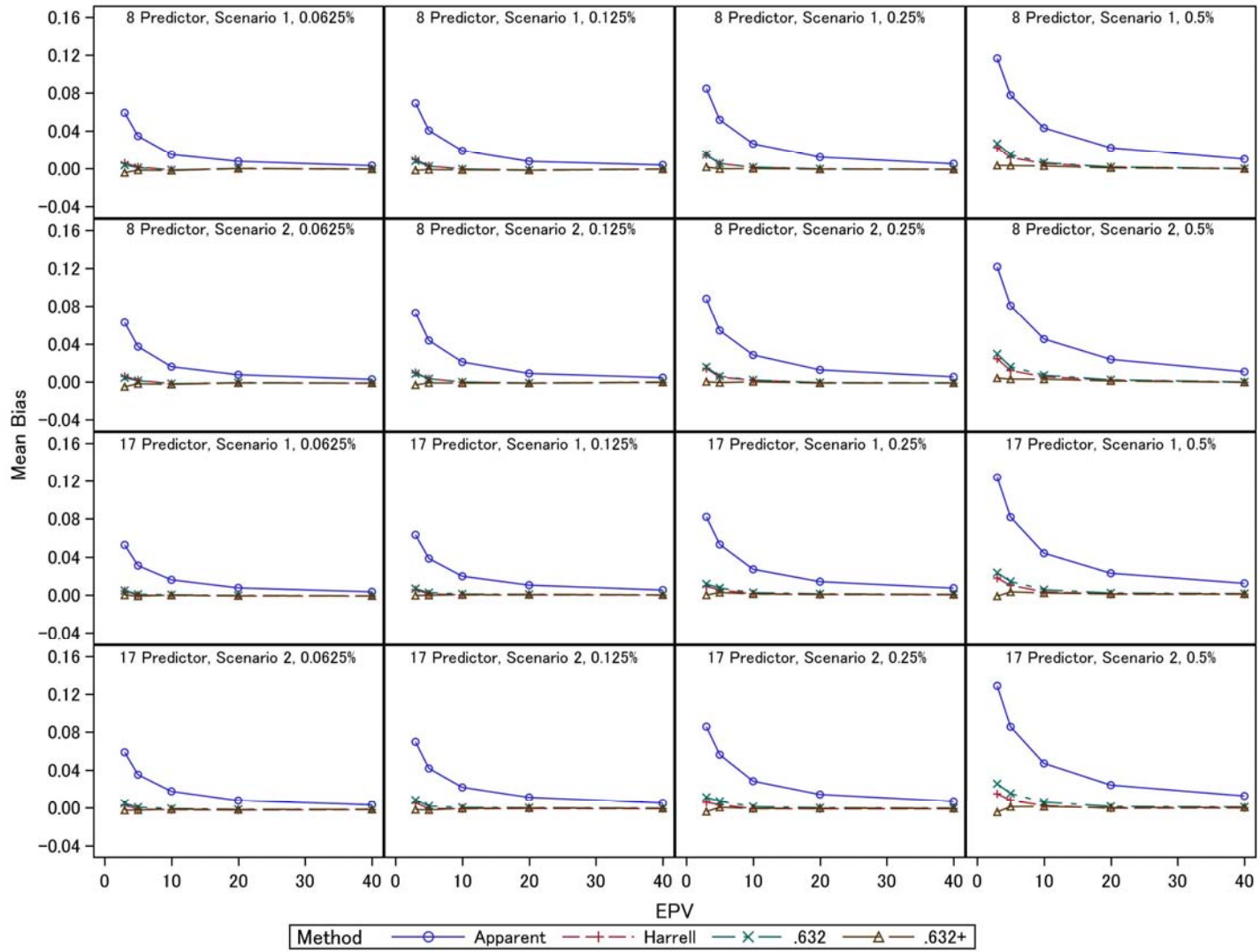

**e-Figure 17.** Simulation results: bias in apparent and optimism corrected *C*-statistics (Stepwise selection; AIC)

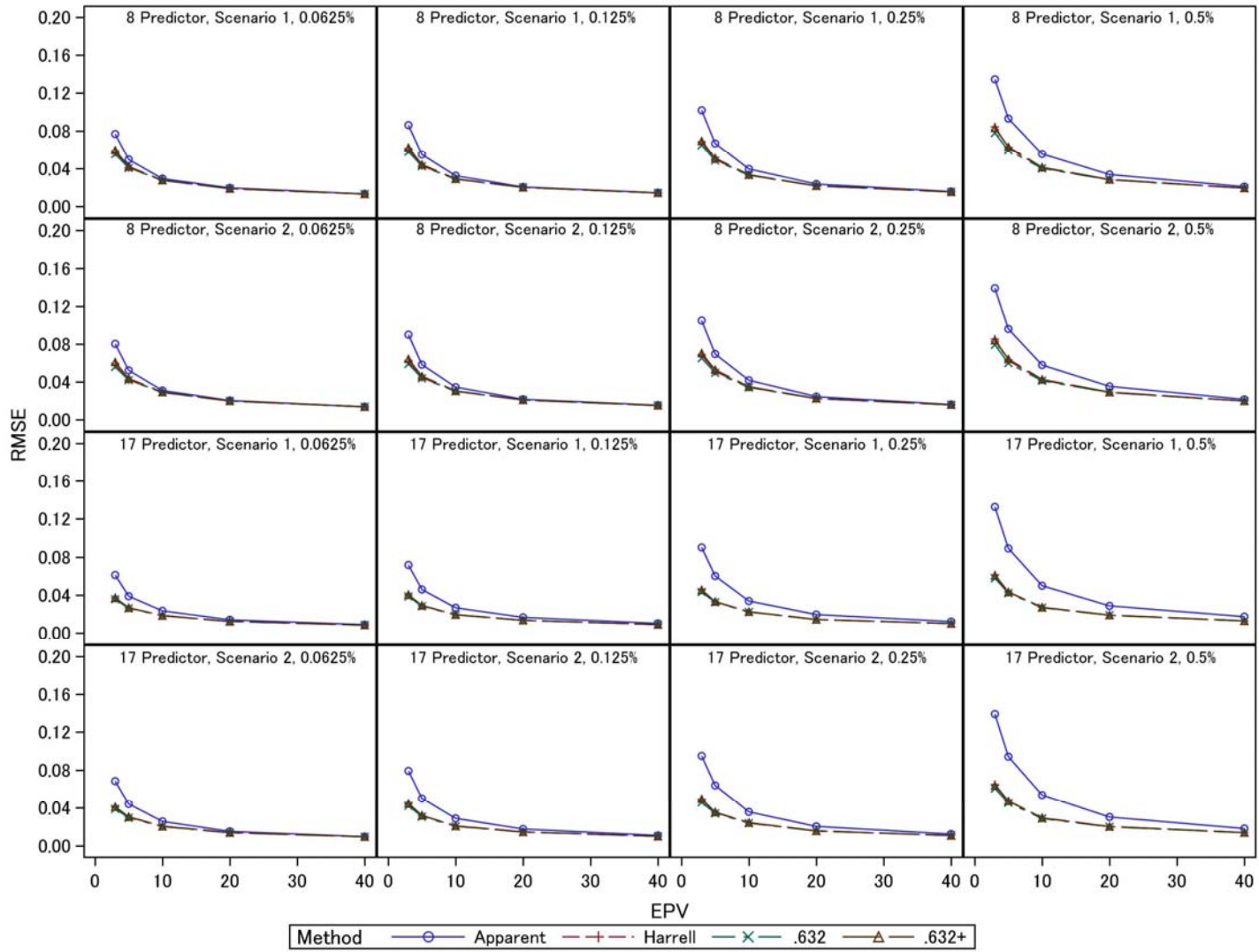

**e-Figure 18.** Simulation results: RMSE in apparent and optimism corrected *C*-statistics (Stepwise selection; AIC)

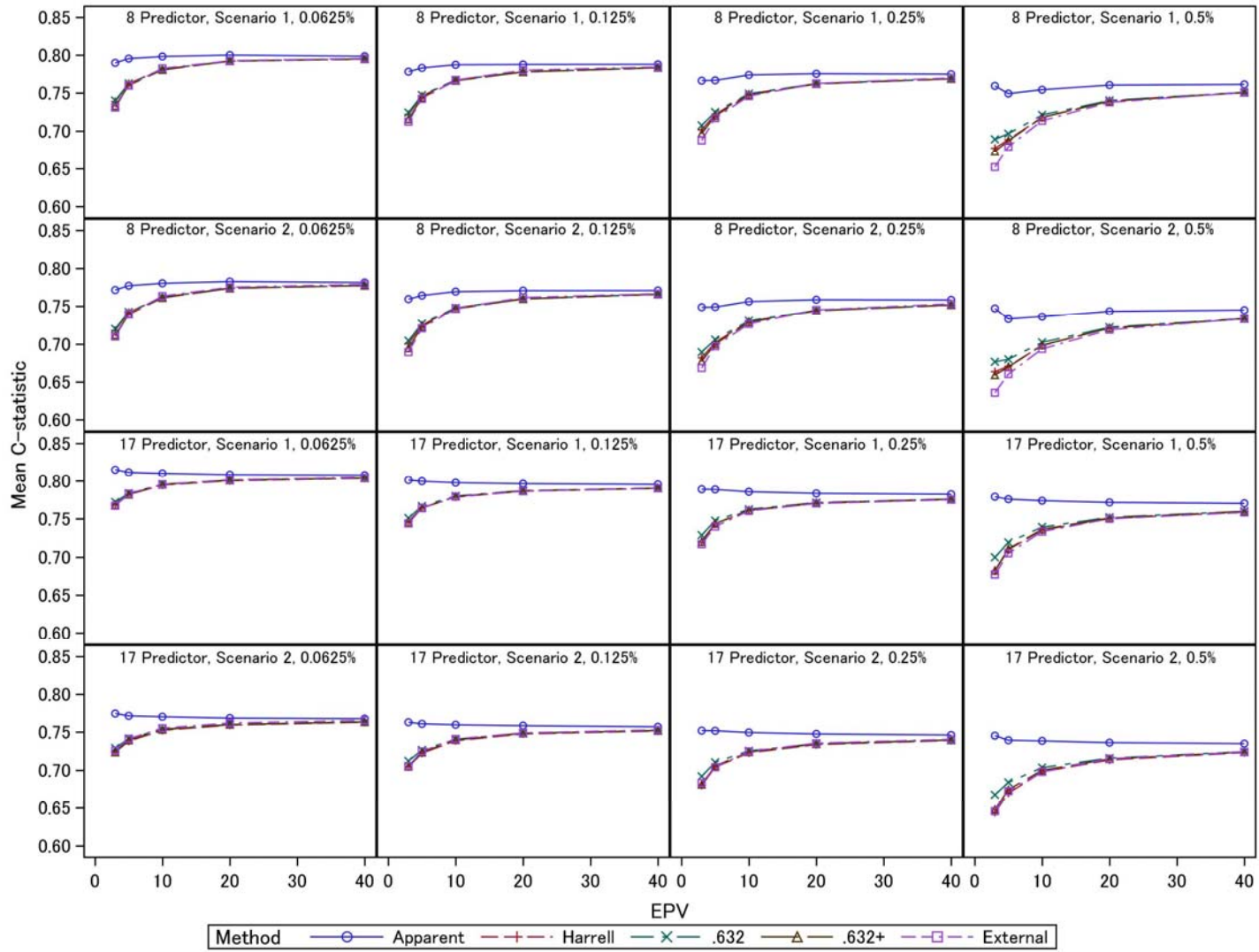

**e-Figure 19.** Simulation results: apparent, external, and optimism corrected *C*-statistics (stepwise selection; P < 0.05)

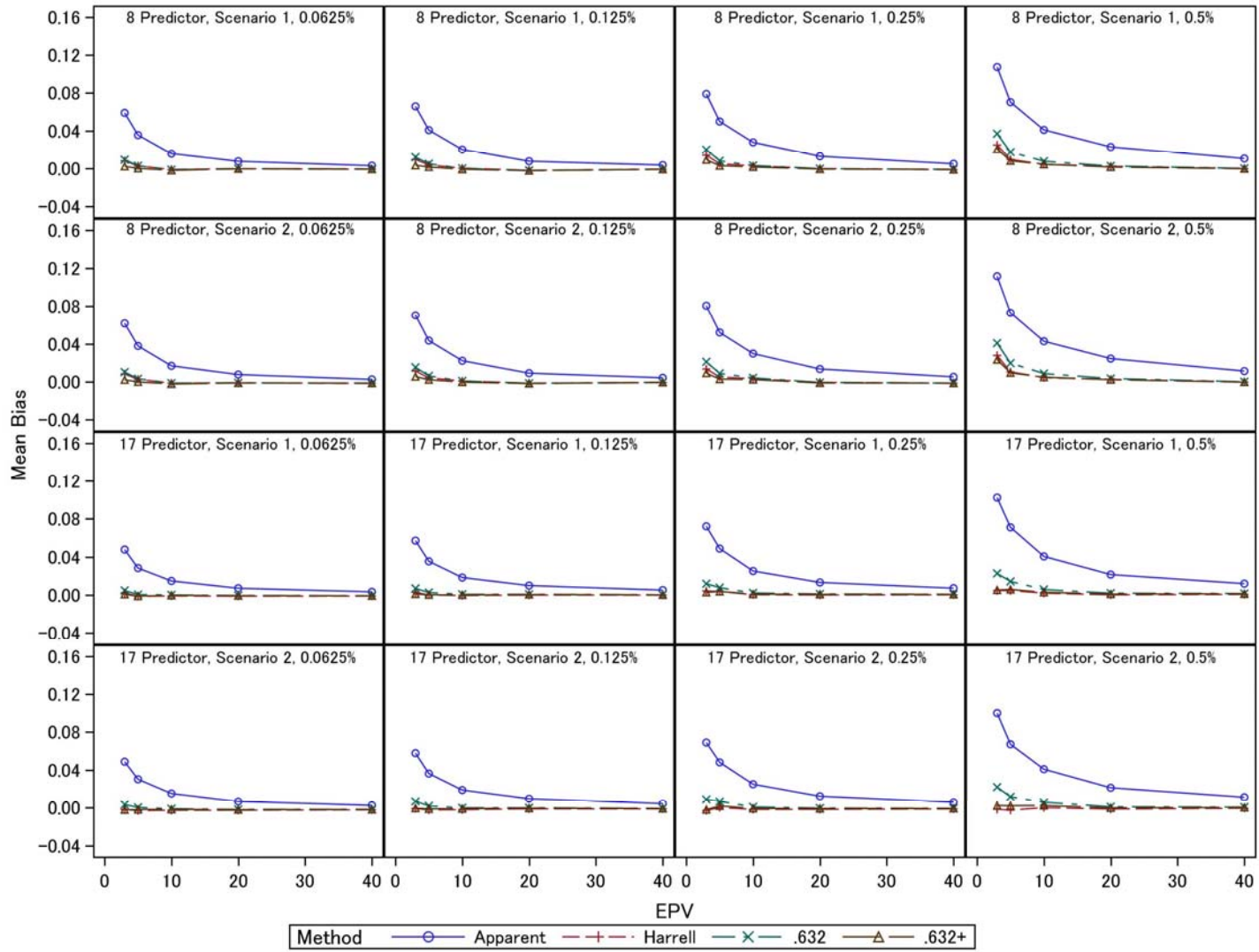

**e-Figure 20.** Simulation results: bias in apparent and optimism corrected *C*-statistics (stepwise selection; P < 0.05)

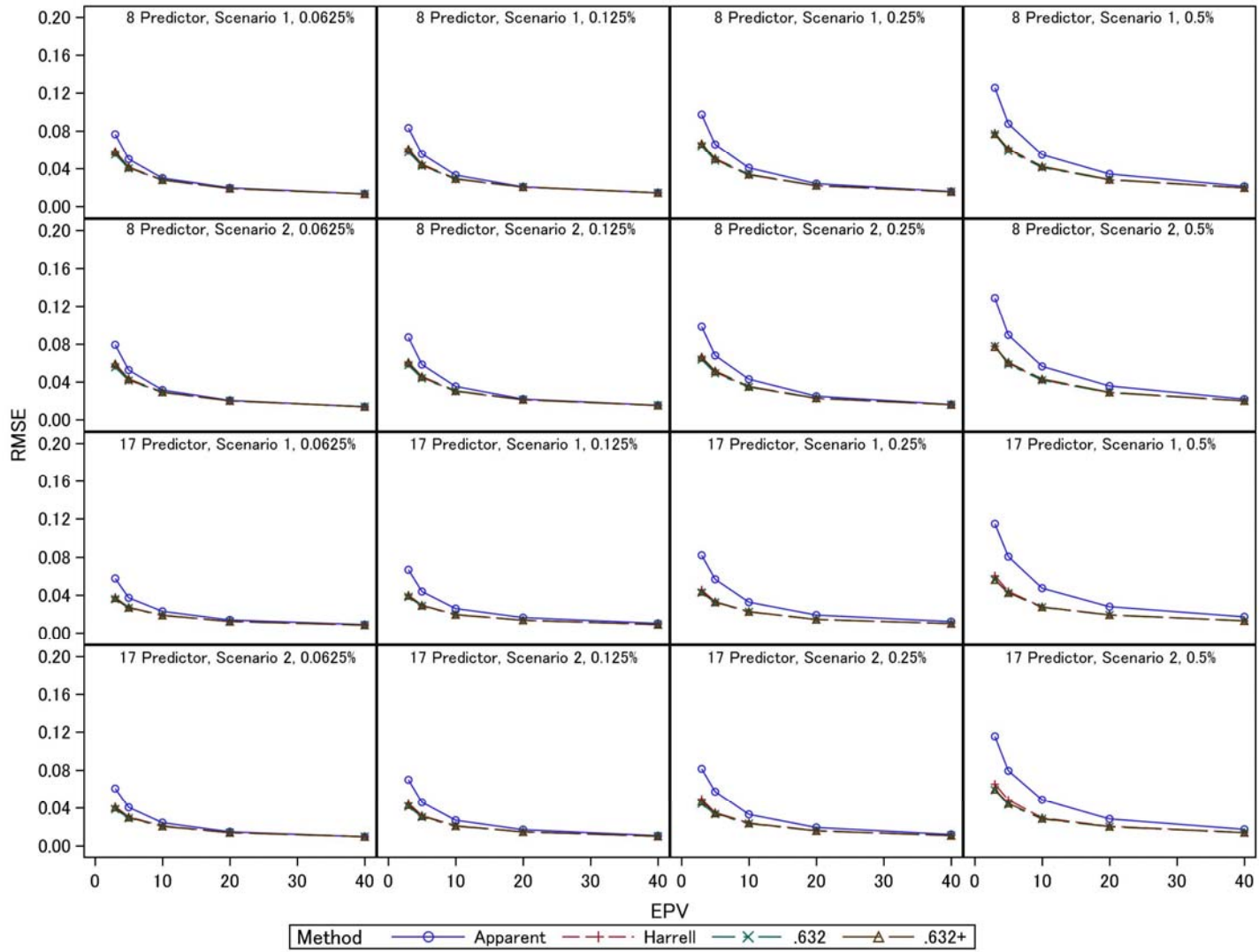

**e-Figure 21.** Simulation results: RMSE in apparent and optimism corrected *C*-statistics (stepwise selection; P < 0.05)

**e-Appendix B: Supplementary tables**

In the Simulations section, we mentioned the final estimated models by lasso, elastic-net, and stepwise selections degenerated to only intercept models at certain frequencies. We present the actual proportions that the only intercept models were obtained from these variable selection methods in e-Table 1.

In addition, we discussed instabilities of the *C*-statistic estimators of lasso predictive models under small sample settings. It can be caused that the tuning parameter was selected by 10-fold CV, i.e., the small datasets were splitted to 10 subgroups, and the resultant individual training datasets can involve only small number of events. To assess the sensitivity of this estimating method, we conducted a supplementary simulations using leave-one-out CV. e-Table 2 show the supplementary results by lasso of the eight-predictor model at EPV = 3.

**e-Table 1.** Proportions (%) that the only intercept models obtained from the variable selection methods: lasso, elastic-net, stepwise selections [†]

| | | Lasso | | | | Elastic-net | | | | Stepwise selection (AIC) | | | | Stepwise selection (p<0.05) | | | |
| | | 8 Predictor | | 17 Predictor | | 8 Predictor | | 17 Predictor | | 8 Predictor | | 17 Predictor | | 8 Predictor | | 17 Predictor | |
| | Events | | | | | | | | | | | | | | | | |
| EPV | fraction | S1 | S2 | S1 | S2 | S1 | S2 | S1 | S2 | S1 | S2 | S1 | S2 | S1 | S2 | S1 | S2 |
|---|---|---|---|---|---|---|---|---|---|---|---|---|---|---|---|---|---|
| 3 | 0.5 | 17.95 | 20.50 | 2.45 | 5.60 | 9.80 | 12.60 | 1.00 | 2.05 | 1.15 | 1.85 | 0.00 | 0.05 | 11.20 | 14.10 | 0.15 | 0.65 |
| 3 | 0.25 | 5.35 | 7.60 | 0.00 | 0.40 | 2.90 | 4.20 | 0.00 | 0.10 | 0.30 | 0.55 | 0.00 | 0.00 | 2.50 | 4.20 | 0.00 | 0.05 |
| 3 | 0.125 | 1.50 | 3.00 | 0.00 | 0.00 | 0.50 | 1.25 | 0.00 | 0.00 | 0.00 | 0.05 | 0.00 | 0.00 | 0.40 | 1.10 | 0.00 | 0.00 |
| 3 | 0.0625 | 1.05 | 1.30 | 0.00 | 0.10 | 0.40 | 0.35 | 0.00 | 0.00 | 0.05 | 0.15 | 0.00 | 0.00 | 0.25 | 0.65 | 0.00 | 0.00 |
| 5 | 0.5 | 3.80 | 5.70 | 0.15 | 0.50 | 1.90 | 3.10 | 0.00 | 0.00 | 0.15 | 0.30 | 0.00 | 0.00 | 2.25 | 3.80 | 0.00 | 0.00 |
| 5 | 0.25 | 0.85 | 1.35 | 0.00 | 0.00 | 0.40 | 0.90 | 0.00 | 0.00 | 0.00 | 0.00 | 0.00 | 0.00 | 0.20 | 0.45 | 0.00 | 0.00 |
| 5 | 0.125 | 0.25 | 0.55 | 0.00 | 0.00 | 0.15 | 0.25 | 0.00 | 0.00 | 0.00 | 0.00 | 0.00 | 0.00 | 0.05 | 0.10 | 0.00 | 0.00 |
| 5 | 0.0625 | 0.00 | 0.10 | 0.00 | 0.00 | 0.00 | 0.05 | 0.00 | 0.00 | 0.00 | 0.00 | 0.00 | 0.00 | 0.00 | 0.00 | 0.00 | 0.00 |
| 10 | 0.5 | 0.10 | 0.15 | 0.00 | 0.00 | 0.00 | 0.00 | 0.00 | 0.00 | 0.00 | 0.00 | 0.00 | 0.00 | 0.00 | 0.05 | 0.00 | 0.00 |
| 10 | 0.25 | 0.00 | 0.00 | 0.00 | 0.00 | 0.00 | 0.00 | 0.00 | 0.00 | 0.00 | 0.00 | 0.00 | 0.00 | 0.00 | 0.00 | 0.00 | 0.00 |
| 10 | 0.125 | 0.00 | 0.00 | 0.00 | 0.00 | 0.00 | 0.00 | 0.00 | 0.00 | 0.00 | 0.00 | 0.00 | 0.00 | 0.00 | 0.00 | 0.00 | 0.00 |
| 10 | 0.0625 | 0.00 | 0.00 | 0.00 | 0.00 | 0.00 | 0.00 | 0.00 | 0.00 | 0.00 | 0.00 | 0.00 | 0.00 | 0.00 | 0.00 | 0.00 | 0.00 |

There were no intercept model at EPV = 20 and EPV = 40.
[†] S1: Scenario 1, S2: Scenario 2

**e-Table 2.** RMSE in apparent and optimism corrected *C*-statistics for lasso of the eight-predictor model based on 10-fold and leave-one-out CV at EPV = 3

| Scenario | Events fraction | 10-fold CV | | | | Leave-one-out CV | | | |
|---|---|---|---|---|---|---|---|---|---|
| | | Apparent | Harrell | .632 | .632+ | Apparent | Harrell | .632 | .632+ |
| 1 | 0.0625 | 0.070 | 0.047 | 0.051 | 0.062 | 0.072 | 0.054 | 0.054 | 0.058 |
| 1 | 0.125 | 0.080 | 0.050 | 0.052 | 0.065 | 0.080 | 0.055 | 0.055 | 0.060 |
| 1 | 0.25 | 0.102 | 0.060 | 0.061 | 0.076 | 0.096 | 0.064 | 0.062 | 0.067 |
| 1 | 0.5 | 0.136 | 0.077 | 0.076 | 0.094 | 0.136 | 0.084 | 0.080 | 0.082 |
| 2 | 0.0625 | 0.075 | 0.048 | 0.054 | 0.067 | 0.076 | 0.054 | 0.055 | 0.059 |
| 2 | 0.125 | 0.085 | 0.050 | 0.054 | 0.069 | 0.086 | 0.057 | 0.057 | 0.061 |
| 2 | 0.25 | 0.106 | 0.061 | 0.062 | 0.079 | 0.101 | 0.065 | 0.063 | 0.067 |
| 2 | 0.5 | 0.145 | 0.080 | 0.079 | 0.096 | 0.142 | 0.086 | 0.082 | 0.083 |